\documentclass[aps,prd,twocolumn,superscriptaddress,showpacs,preprintnumbers]{revtex4-1}

\usepackage{graphicx}
\usepackage{amsmath}
\usepackage{amsfonts}
\usepackage{multirow}
\usepackage{xcolor}

\newcommand{\B}{{\bar{B}^0}}
\newcommand{\jp}{{J/\psi}}
\newcommand{\psp}{{\psi(2S)}}

\newcommand{\km}{{K^-}}
\newcommand{\ks}{{K^0_S}}
\newcommand{\pip}{{\pi^+}}
\newcommand{\pim}{{\pi^-}}
\newcommand{\z}{{Z_c^+}}
\newcommand{\zp}[1]{{Z_c(#1)^+}}

\newcommand{\zpm}[1]{{Z_c(#1)^\pm}}
\newcommand{\zs}{{Z_{cs}^-}}
\newcommand{\kst}{{K^*}}
\newcommand{\lpair}{{\ell^+\ell^-}}
\newcommand{\lp}{{\ell^+}}
\newcommand{\lm}{{\ell^-}}
\newcommand{\elp}{{e^+}}
\newcommand{\elm}{{e^-}}
\newcommand{\mup}{{\mu^+}}
\newcommand{\mum}{{\mu^-}}
\newcommand{\ee}{{e^+e^-}}
\newcommand{\mumu}{{\mu^+\mu^-}}
\newcommand{\decay}{\B \to \jp \km \pip}
\newcommand{\decayll}{\B \to \jp (\to \lp \lm) \km \pip}

\newcommand{\fb}{\mathrm{fb}^{-1}}
\newcommand{\cm}{\mathrm{cm}}
\newcommand{\mev}{\mathrm{MeV}}
\newcommand{\mevcc}{\mathrm{MeV}/c^2}

\newcommand{\gevcc}{\mathrm{GeV}/c^2}
\newcommand{\gevccsq}{\mathrm{GeV}^2/c^4}
\newcommand{\mbc}{M_{\mathrm{bc}}}
\newcommand{\anorm}{{A_\bot}}
\newcommand{\atang}{{A_{\|}}}
\newcommand{\ebeam}{E_{\mathrm{beam}}}
\newcommand{\sx}{M^2_{K \pi}}
\newcommand{\sy}{{M^2_{\jp \pi}}}
\newcommand{\DE}{\Delta E}
\newcommand{\br}{\mathcal{B}}
\newcommand{\dlnl}{{\Delta(-2\ln\mathit{L})}}

\begin{document}


\title{Observation of a new charged charmoniumlike state in $\decay$ decays}

\noaffiliation
\affiliation{University of the Basque Country UPV/EHU, 48080 Bilbao}
\affiliation{Beihang University, Beijing 100191}
\affiliation{University of Bonn, 53115 Bonn}
\affiliation{Budker Institute of Nuclear Physics SB RAS and Novosibirsk State University, Novosibirsk 630090}
\affiliation{Faculty of Mathematics and Physics, Charles University, 121 16 Prague}
\affiliation{Chonnam National University, Kwangju 660-701}
\affiliation{University of Cincinnati, Cincinnati, Ohio 45221}
\affiliation{Deutsches Elektronen--Synchrotron, 22607 Hamburg}
\affiliation{Justus-Liebig-Universit\"at Gie\ss{}en, 35392 Gie\ss{}en}
\affiliation{The Graduate University for Advanced Studies, Hayama 240-0193}
\affiliation{Gyeongsang National University, Chinju 660-701}
\affiliation{Hanyang University, Seoul 133-791}
\affiliation{University of Hawaii, Honolulu, Hawaii 96822}
\affiliation{High Energy Accelerator Research Organization (KEK), Tsukuba 305-0801}
\affiliation{IKERBASQUE, Basque Foundation for Science, 48011 Bilbao}
\affiliation{Indian Institute of Technology Guwahati, Assam 781039}
\affiliation{Indian Institute of Technology Madras, Chennai 600036}
\affiliation{Institute of High Energy Physics, Chinese Academy of Sciences, Beijing 100049}
\affiliation{Institute of High Energy Physics, Vienna 1050}
\affiliation{Institute for High Energy Physics, Protvino 142281}
\affiliation{INFN - Sezione di Torino, 10125 Torino}
\affiliation{Institute for Theoretical and Experimental Physics, Moscow 117218}
\affiliation{J. Stefan Institute, 1000 Ljubljana}
\affiliation{Kanagawa University, Yokohama 221-8686}
\affiliation{Institut f\"ur Experimentelle Kernphysik, Karlsruher Institut f\"ur Technologie, 76131 Karlsruhe}
\affiliation{Department of Physics, Faculty of Science, King Abdulaziz University, Jeddah 21589}
\affiliation{Korea Institute of Science and Technology Information, Daejeon 305-806}
\affiliation{Korea University, Seoul 136-713}
\affiliation{Kyungpook National University, Daegu 702-701}
\affiliation{\'Ecole Polytechnique F\'ed\'erale de Lausanne (EPFL), Lausanne 1015}
\affiliation{Faculty of Mathematics and Physics, University of Ljubljana, 1000 Ljubljana}
\affiliation{Luther College, Decorah, Iowa 52101}
\affiliation{University of Maribor, 2000 Maribor}
\affiliation{Max-Planck-Institut f\"ur Physik, 80805 M\"unchen}
\affiliation{School of Physics, University of Melbourne, Victoria 3010}
\affiliation{Moscow Physical Engineering Institute, Moscow 115409}
\affiliation{Moscow Institute of Physics and Technology, Moscow Region 141700}
\affiliation{Graduate School of Science, Nagoya University, Nagoya 464-8602}
\affiliation{Kobayashi-Maskawa Institute, Nagoya University, Nagoya 464-8602}
\affiliation{Nara Women's University, Nara 630-8506}
\affiliation{National Central University, Chung-li 32054}
\affiliation{National United University, Miao Li 36003}
\affiliation{Department of Physics, National Taiwan University, Taipei 10617}
\affiliation{H. Niewodniczanski Institute of Nuclear Physics, Krakow 31-342}
\affiliation{Niigata University, Niigata 950-2181}
\affiliation{Osaka City University, Osaka 558-8585}
\affiliation{Pacific Northwest National Laboratory, Richland, Washington 99352}
\affiliation{Peking University, Beijing 100871}
\affiliation{University of Pittsburgh, Pittsburgh, Pennsylvania 15260}
\affiliation{University of Science and Technology of China, Hefei 230026}
\affiliation{Seoul National University, Seoul 151-742}
\affiliation{Soongsil University, Seoul 156-743}
\affiliation{Sungkyunkwan University, Suwon 440-746}
\affiliation{School of Physics, University of Sydney, NSW 2006}
\affiliation{Department of Physics, Faculty of Science, University of Tabuk, Tabuk 71451}
\affiliation{Tata Institute of Fundamental Research, Mumbai 400005}
\affiliation{Excellence Cluster Universe, Technische Universit\"at M\"unchen, 85748 Garching}
\affiliation{Toho University, Funabashi 274-8510}
\affiliation{Tohoku University, Sendai 980-8578}
\affiliation{Department of Physics, University of Tokyo, Tokyo 113-0033}
\affiliation{Tokyo Institute of Technology, Tokyo 152-8550}
\affiliation{Tokyo Metropolitan University, Tokyo 192-0397}
\affiliation{University of Torino, 10124 Torino}
\affiliation{CNP, Virginia Polytechnic Institute and State University, Blacksburg, Virginia 24061}
\affiliation{Wayne State University, Detroit, Michigan 48202}
\affiliation{Yonsei University, Seoul 120-749}
  \author{K.~Chilikin}\affiliation{Institute for Theoretical and Experimental Physics, Moscow 117218} 
  \author{R.~Mizuk}\affiliation{Institute for Theoretical and Experimental Physics, Moscow 117218}\affiliation{Moscow Physical Engineering Institute, Moscow 115409} 
  \author{I.~Adachi}\affiliation{High Energy Accelerator Research Organization (KEK), Tsukuba 305-0801}\affiliation{The Graduate University for Advanced Studies, Hayama 240-0193} 
  \author{H.~Aihara}\affiliation{Department of Physics, University of Tokyo, Tokyo 113-0033} 
  \author{S.~Al~Said}\affiliation{Department of Physics, Faculty of Science, University of Tabuk, Tabuk 71451}\affiliation{Department of Physics, Faculty of Science, King Abdulaziz University, Jeddah 21589} 
  \author{K.~Arinstein}\affiliation{Budker Institute of Nuclear Physics SB RAS and Novosibirsk State University, Novosibirsk 630090} 
  \author{D.~M.~Asner}\affiliation{Pacific Northwest National Laboratory, Richland, Washington 99352} 
  \author{V.~Aulchenko}\affiliation{Budker Institute of Nuclear Physics SB RAS and Novosibirsk State University, Novosibirsk 630090} 
  \author{T.~Aushev}\affiliation{Institute for Theoretical and Experimental Physics, Moscow 117218} 
  \author{R.~Ayad}\affiliation{Department of Physics, Faculty of Science, University of Tabuk, Tabuk 71451} 
% ...
\collaboration{The Belle Collaboration}\noaffiliation

\begin{abstract}
We present the results of an amplitude analysis of $\decay$ decays.
A new charged charmoniumlike state $\zp{4200}$ decaying to $\jp\pip$ is
observed with a significance of $6.2\sigma$.
The mass and width of the $\zp{4200}$
are $4196^{+31 +17}_{-29 -13}\ \mevcc$
and $370^{+70 +70}_{-70 -132}\ \mev$,
respectively; the preferred assignment of the quantum numbers is $J^P=1^+$.
In addition, we find evidence for $\zp{4430}\to\jp\pip$.
The analysis is based on a 711 $\fb$ data sample collected by the
Belle detector at the asymmetric-energy $\ee$ collider KEKB.
\end{abstract}


\pacs{14.40.Nd, 14.40.Rt, 13.25.-k}


\maketitle


\section{Introduction}

Recently, a number of new states containing a $c\bar{c}$ quark pair have been
observed, many of which are not well described by
the quark model~\cite{olsen_godfrey, brambilla, brambilla_qcd_review}.
Among these states are charged charmoniumlike state candidates with a
minimal quark content that is necessarily exotic:
$|c\bar{c}u\bar{d}\rangle$.
The first of these states, the $\zp{4430}$, was observed
by the Belle Collaboration in the $\psp\pip$ invariant mass spectrum in
$\bar{B}^0\to\psp\km\pip$ decays~\cite{choiolsen, z4430dalitz, z4430jp}.
Two other states, the $\zp{4050}$ and $\zp{4250}$,
were observed by Belle in the $\chi_{c1}\pip$ invariant mass spectrum in
$\bar{B}^0\to\chi_{c1}\km\pip$ decays~\cite{mizukchistov}.
The BaBar Collaboration searched for these states
\cite{babarjpkpi, babarchickpi} but did not confirm them.
However, recently, the LHCb collaboration confirmed the Belle observation of
the $\zp{4430}$ with overwhelming ($>14\sigma$) significance~\cite{z4430lhcb}.
The BESIII and Belle Collaborations observed
the $\zpm{3900}$ in the $J/\psi \pi^\pm$
invariant mass spectrum in $Y(4260)\to J/\psi \pi^+ \pi^-$
decays~\cite{z3900bes, z3900belle}.
The charged $\zpm{3900}$ was also
observed in CLEO data~\cite{z3900cleo}; in this analysis, evidence for
the neutral $Z_c(3900)^0$ was also found.
The $\zpm{3885}$, which is likely to be the same state,
was observed by the BESIII
Collaboration in $e^+ e^- \to (D \bar{D}^*)^\pm \pi^\mp$~\cite{z3900bes_ddst}.
Also, the BESIII Collaboration
observed the $\zpm{4020}$ in the $h_c \pi^\pm$ invariant mass spectrum
in $e^+ e^- \to h_c \pip \pim$~\cite{z4020}.
Finally, the $\zpm{4025}$ was observed by the BESIII Collaboration in
$e^+ e^- \to (D^* \bar{D}^*)^\pm \pi^\mp$~\cite{bes_dstdst}.

Here we present the results of a full amplitude analysis of the decay $\decay$,
with $\jp \to \mu^+\mu^-$ or $\jp \to e^+ e^-$.
The analysis is similar to the Belle study of
$\bar{B}^0 \to \psp K^- \pi^+$~\cite{z4430jp}.
It is performed using a $711\ \fb$ data sample
collected by the Belle detector
at the KEKB asymmetric-energy $\ee$ collider~\cite{kekb}.
The data sample was collected at the
$\Upsilon(4S)$ resonance and contains $772\times10^6$ $B\bar{B}$ pairs.


\section{The Belle Detector}

The Belle detector is a large-solid-angle magnetic
spectrometer that consists of a silicon vertex detector (SVD),
a 50-layer central drift chamber (CDC), an array of
aerogel threshold Cherenkov counters (ACC),
a barrel-like arrangement of time-of-flight
scintillation counters (TOF), and an electromagnetic calorimeter
comprised of CsI(Tl) crystals (ECL) located inside 
a superconducting solenoid coil that provides a 1.5~T
magnetic field.  An iron flux-return located outside of
the coil is instrumented to detect $K_L^0$ mesons and to identify
muons (KLM).  The detector
is described in detail elsewhere~\cite{Belle}.
Two inner detector configurations were used. A 2.0 cm beampipe
and a 3-layer silicon vertex detector were used for the first sample
of 140 $\fb$, while a 1.5 cm beampipe, a 4-layer
silicon detector and a small-cell inner drift chamber were used to record  
the remaining 571 $\fb$\cite{svd2}.  

We use a GEANT-based Monte Carlo (MC) simulation~\cite{geant} to model
the response of the detector, identify potential backgrounds and
determine the acceptance. The MC simulation includes run-dependent
detector performance variations and background conditions.  Signal MC
events are generated with EvtGen~\cite{evtgen}
in proportion to the relative luminosities of the
different running periods.


\section{Event selection}

We select events of the type $\decay$ (where inclusion of charge-conjugate
modes is always implied), with the $\jp$ meson reconstructed via its
$\ee$ and $\mumu$ decay channels.
The selection procedure is identical
to that in Ref.~\cite{z4430jp} with the replacement of the $\psp$ by the $\jp$.

All tracks are required to originate from the interaction region,
$dr < 0.2\ \cm$ and $|dz| < 2\ \cm$, where $dr$ and $dz$ are
the cylindrical coordinates (the radial distance and longitudinal position,
respectively, with
the $z$ axis of the reference frame antiparallel to the positron beam axis
and the origin being the run-dependent mean interaction point)
of the point of closest approach of the track to
the $z$ axis in the interaction region.
Charged $\pi$ and $K$ mesons are identified using
likelihood ratios $R_{\pi/K} = \mathcal{L}_\pi/(\mathcal{L}_\pi+\mathcal{L}_K)$
and $R_{K/\pi} = \mathcal{L}_K/(\mathcal{L}_\pi+\mathcal{L}_K)$,
where $\mathcal{L}_\pi$ and $\mathcal{L}_K$ are
likelihoods, respectively, for $\pi$ and $K$. The likelihoods
 are calculated from the
time-of-flight information from the TOF, the number of photoelectrons from
the ACC and $dE/dx$ measurements in the CDC. We require $R_{\pi/K}>0.6$
for $\pi$ candidates and $R_{K/\pi}>0.6$ for $K$ candidates.
The $K$ identification efficiency is typically 90\% and
the misidentification probability is about 10\%. Muons are identified by
their range and transverse scattering in the KLM.
Electrons are identified by the presence of a matching
electromagnetic shower in the ECL. An electron
veto is imposed on $\pi$ and $K$ candidates.

For $\jp\to\ee$ candidates, we collect bremsstrahlung radiation by
including photons that have energies greater
than 30 $\mev$ and are within 50 mrad of the lepton direction
in the calculation of the $\jp$ invariant mass.
We require $|M(\lpair)-m_{\jp}|<60\ \mevcc$, where $\ell$
is either $\mu$ or $e$.
We perform a mass-constrained fit to the $\jp$ candidates.
The data from $\ee$ and $\mumu$ channels are combined since
both channels have the same angular distribution.

The beam-energy-constrained mass of the $B$ meson is defined as
$\mbc = \sqrt{E_{\mathrm{beam}}^2-(\sum_i\vec{p}_i)^2}$,
where $E_{\mathrm{beam}}$ is the
beam energy in the center-of-mass frame and $\vec{p}_i$ are the momenta
of the decay products in the same frame. We require
$|\mbc-m_B| < 7\ \mevcc$, where $m_B$ is the $B^0$ mass~\cite{PDG}.
A mass-constrained fit is applied to the $B$ meson candidates.

\section{Event distributions and signal yield}

\begin{figure}[ht]
\begin{center}
\includegraphics[width=6cm]{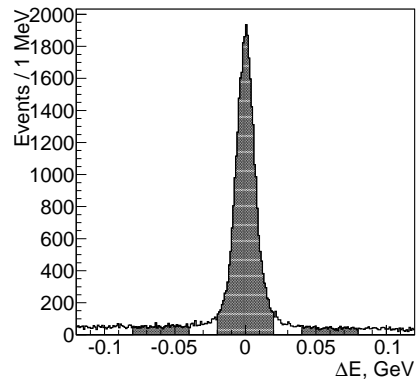}
\end{center}
\caption{The $\DE$ distribution; the signal and sideband regions are hatched.}
\label{fig:deltae}
\end{figure}

The difference between the reconstructed energy and the beam energy
$\DE = \sum_i E_i - \ebeam$, where $E_i$ are energies of the $\B$ decay
products in the center-of-mass frame, is used to identify the signal.
The signal region is defined as $|\DE| < 20\ \mev$, and the sidebands
are defined as $40\ \mev < |\DE| < 80\ \mev$.
The $\DE$ distribution with marked signal and sideband regions
is shown in Fig.~\ref{fig:deltae}. 

To determine the signal and background event yields, we perform a binned
maximum likelihood fit of the $\DE$ distribution that is modeled by
the sum of two Gaussian functions to represent signal and a
second-order polynomial for the background.
The total number of events in the signal region is 31\,774 and
the number of signal events in the signal region is
$29\,990\pm190\pm50$ (here and
elsewhere, the first uncertainty is statistical and the second
is systematic). The systematic error is estimated by changing
the $\DE$ fit interval and the order of the polynomial.

\begin{figure}[ht]
\begin{center}
\includegraphics[width=6cm]{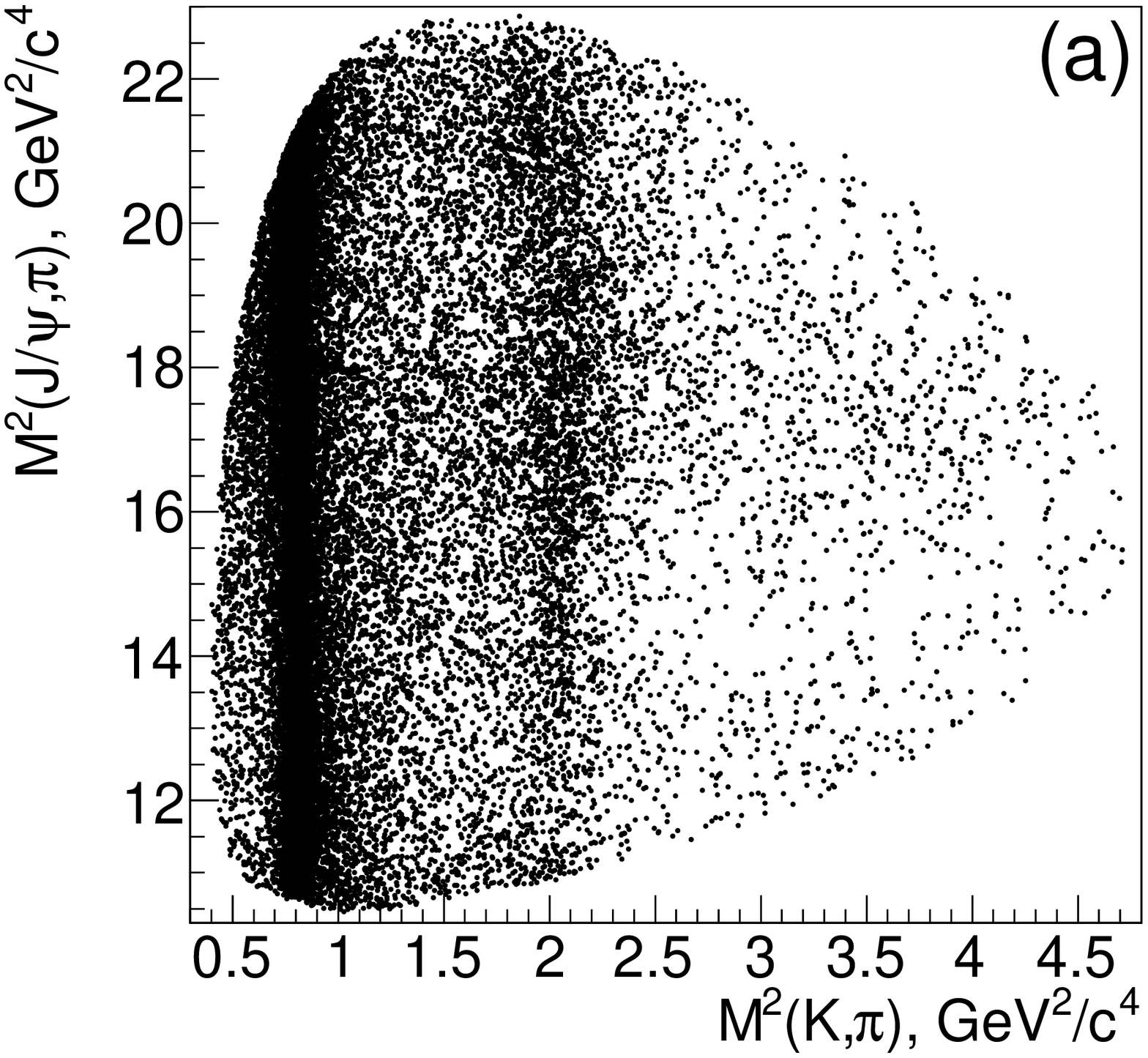}\newline
\includegraphics[width=6cm]{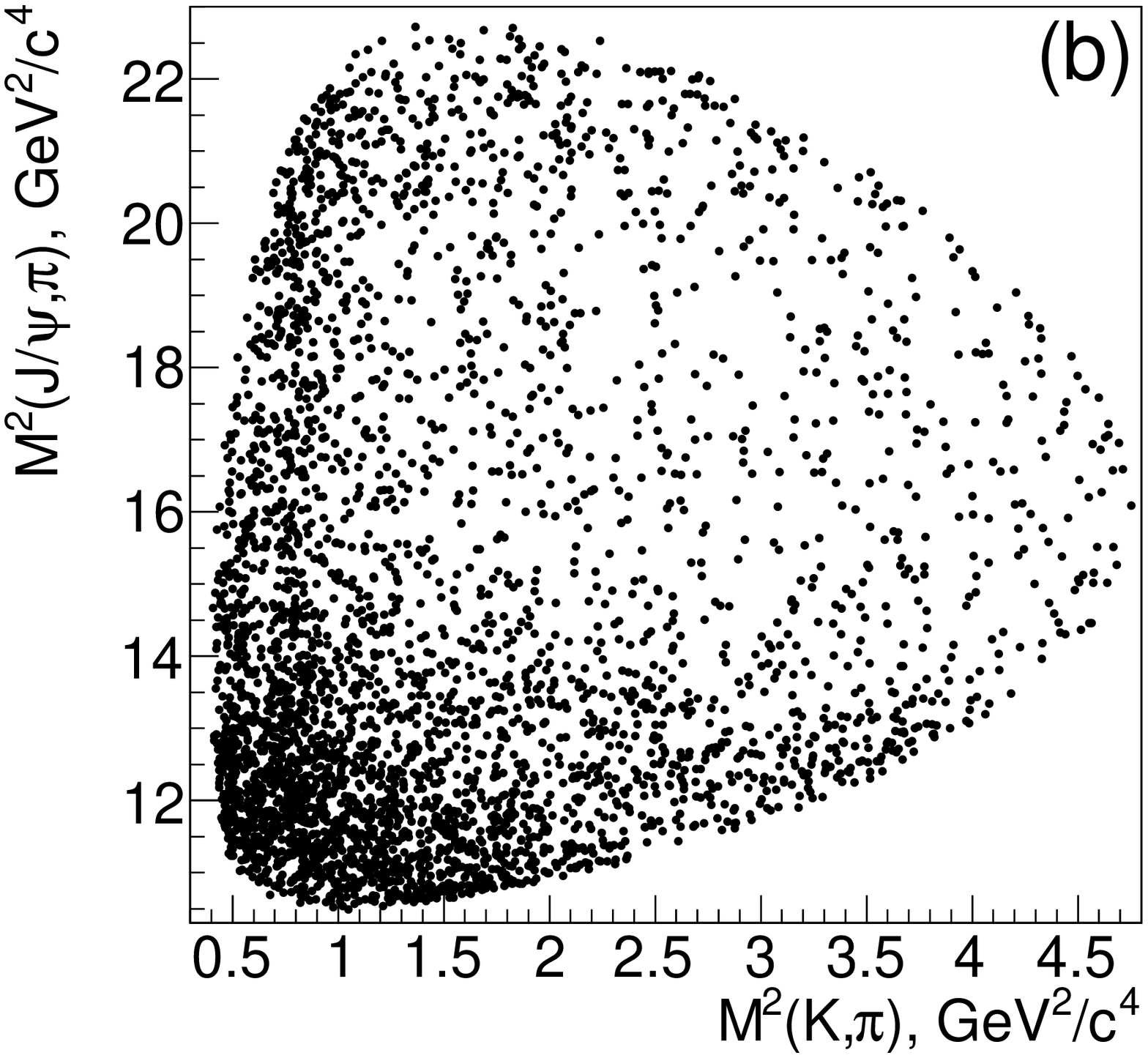}\newline
\includegraphics[width=6cm]{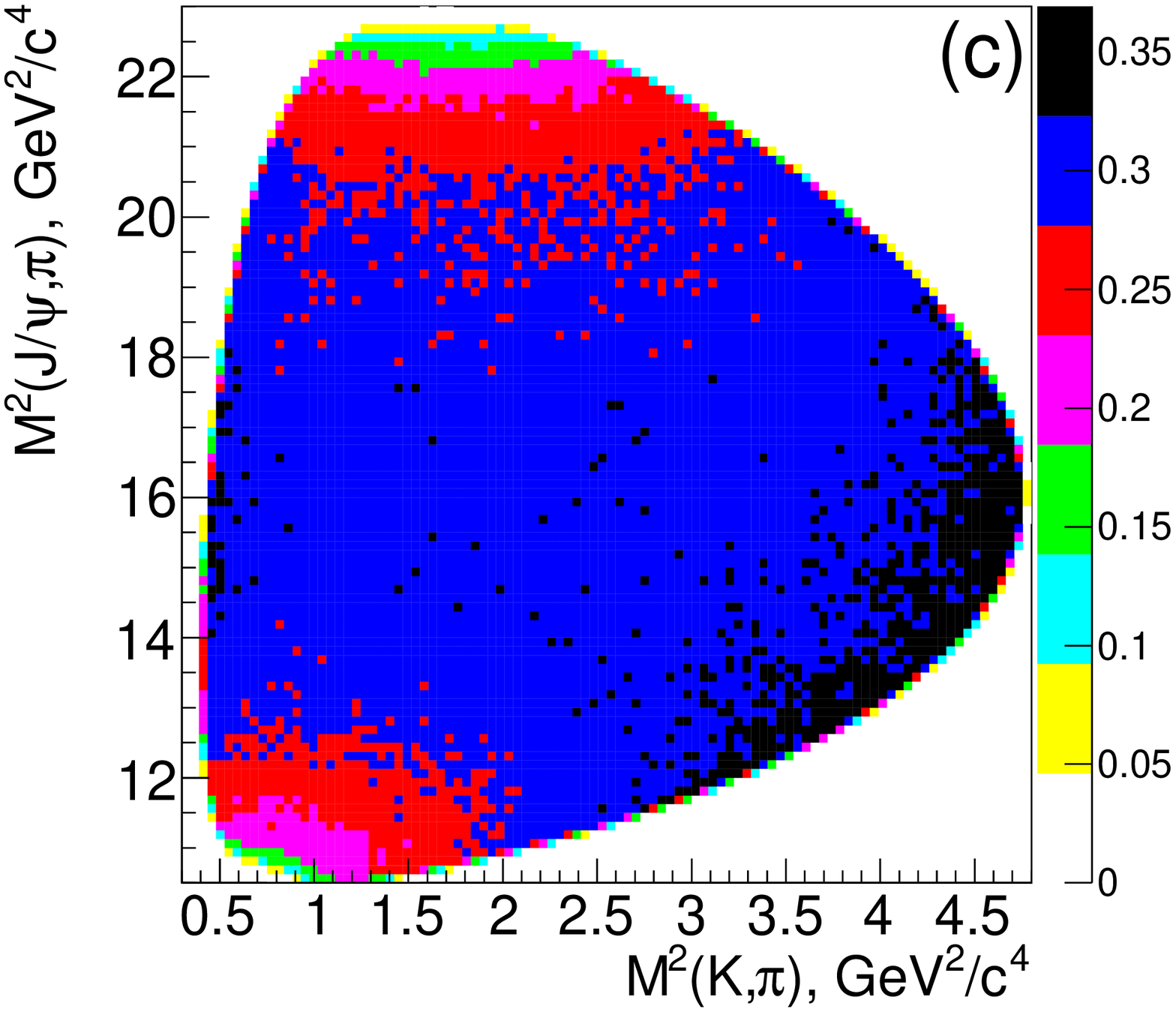}\newline
\end{center}
\caption{Dalitz plots of the signal region (a), sidebands (b) and
signal efficiency (c).}
\label{fig:dalitz}
\end{figure}

The Dalitz plot for the signal
region is shown in Fig.~\ref{fig:dalitz}(a). The most prominent features are
the vertical bands due to the production of intermediate $K^*(892)$ and
$K^*_2(1430)$ resonances. The Dalitz plot for the sidebands
is shown in Fig.~\ref{fig:dalitz}(b), where the events primarily accumulate in
the lower left corner where the momentum of pions is low.

To determine the reconstruction efficiency, we generate MC events for
$\decayll$ with a uniform
phase space distribution. The efficiency is corrected for the difference
between the particle identification efficiency in
data and MC, which is obtained from a $D^{*+}\to D^0(\to K^-\pi^+)\pi^+$
control sample for $K$ and $\pi$ and a sample of $\gamma\gamma\to\lpair$
for $\mu$ and $e$.

\begin{figure}[ht]
\includegraphics[width=4.2cm]{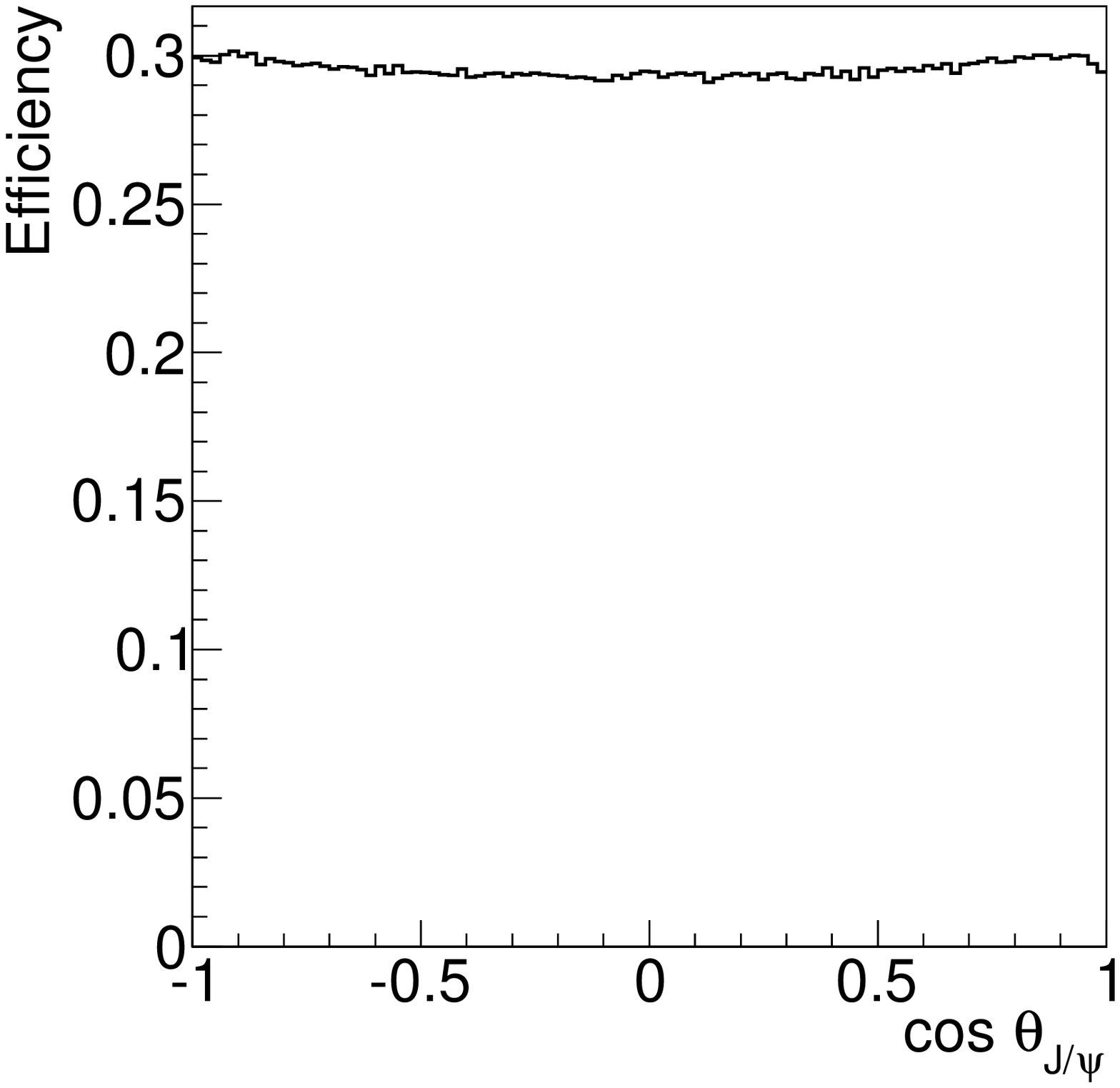}
\includegraphics[width=4.2cm]{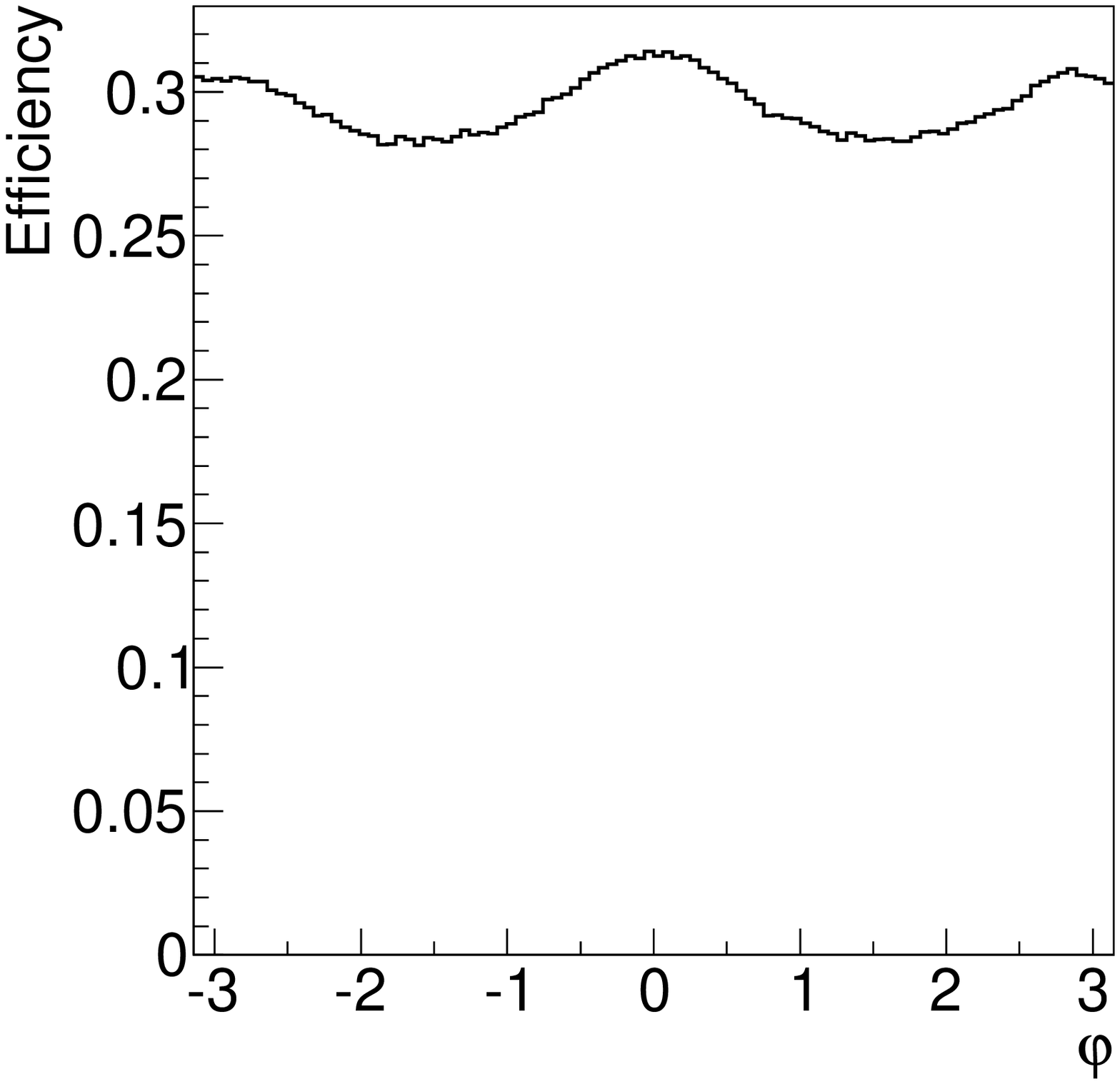}
\caption{Efficiency as a function of the angular variables.}
\label{fig:angeff}
\end{figure}

The efficiency as a function of the Dalitz
variables is shown in Fig.~\ref{fig:dalitz}(c).
The efficiency drops in the
lower left corner where the pions have low momentum and in
the upper corner where the kaons have low momentum;
elsewhere, it is almost uniform.
The efficiency as a function of the angular variables is shown
in Fig.~\ref{fig:angeff}; $\theta_{\jp}$ is the $\jp$ helicity angle,
defined as the angle between the momenta of the $(\km,\pip)$ system and
the $\lm$ in the $\jp$ rest frame, and
$\varphi$ is the angle between the planes defined by
the $(\lp,\lm)$ and $(\km,\pip)$ momenta
in the $\B$ rest frame.
The efficiency is almost independent of $\cos \theta_{\jp}$;
its dependence on $\varphi$ is stronger, with a variation that
is at the 10\% level.

\section{Amplitude analysis formalism}

The amplitude of the decay $\decayll$ is
represented as the sum of Breit-Wigner contributions for different
intermediate two-body states.
The amplitude is calculated using the helicity formalism
in a four-dimensional parameter space, defined as
\begin{equation}
\Phi = (\sx, \sy, \theta_\jp, \varphi).
\end{equation}
The contributions of each individual $\kst$ resonance
and the $\z$ resonance to the signal density function $S(\Phi)$
are the same as in Ref.~\cite{z4430jp}; the definition of the helicity 
amplitudes $H_\lambda$ is also the same.
The difference from Ref.~\cite{z4430jp} is that
the default model includes more $\kst$ resonances due to the
larger accessible kinematic range
(up to $M_{K \pi} = 2183\ \mevcc$).
The known resonances included in the default model
are $K^*_0(800)$, $K^*(892)$, $K^*(1410)$, $K^*_0(1430)$,
$K^*_2(1430)$, $K^*(1680)$, $K^*_3(1780)$, $K^*_0(1950)$, $K^*_2(1980)$,
$K^*_4(2045)$ and $\zp{4430}$; a search for
additional exotic $\z$ resonances is performed.

The background density function is
\begin{equation}
\begin{split}
B(\Phi) = &
 (B_{\mathrm{sm}}(\Phi)+B_\kst(\Phi)+B_{\ks}(\Phi)) \\
& \times P_{\theta_{\jp}}(\cos\theta_{\jp}) P_\varphi(\varphi),
\end{split}
\label{eq:bgpdf}
\end{equation}
where $B_{\mathrm{sm}}$ is the smooth part of the background,
$B_\kst$ is the background from the $\kst(892)$ mesons, $B_{\ks}$ is the
$\ks\to\pi^+\pi^-$ background (where one of the $\pi$ mesons is misidentified
as a $K$) and $P_{\theta_{\jp}}$ and $P_\varphi$ are second-order polynomials.

The smooth part of the background is described by
\begin{equation}
\begin{split}
B_{\mathrm{sm}}(\Phi) = & (\alpha_1 e^{-\beta_1 M^2_{K^- \pi^+}} +
\alpha_2 e^{-\beta_2 M^2_{\jp K^-}}) \\
& \times P_{\mathrm{sm}}(\sx, \sy),
\end{split}
\end{equation}
where $\alpha_1$, $\alpha_2$, $\beta_1$ and $\beta_2$ are real parameters and
$P_{\mathrm{sm}}$ is a two-dimensional fifth-order polynomial.
The background originating from the $\kst(892)$ mesons is described by the
function
\begin{equation}
B_\kst(\Phi) = |A^{\kst(892)}(\sx)|^2 P_\kst(\sy),
\end{equation}
where $A^{\kst(892)}$ is the Breit-Wigner amplitude of the $\kst(892)$ and
$P_\kst$ is a fourth-order polynomial.

Background events from $\ks\to\pi^+\pi^-$ decays
have a specific $\sx$ dependence on $\sy$:
\begin{equation}
\begin{split}
\sx(\ks) = &
 M_{\ks}^2+M_{K^+}^2-M_{\pi^+}^2 \\
+ &
\frac{M_{\ks}^2 + M_{\pi^+}^2 - M_{\jp}^2 + \sy}{M_{B^0}} \\
& \times\left(\sqrt{E_\pi^2 + M_{K^+}^2 - M_\pi^2} -E_\pi\right),
\end{split}
\end{equation}
where
\begin{equation}
E_\pi = \frac{M_{B^0}^2 + M_{\pi^+}^2 - \sy}{2 M_{B^0}}
\end{equation}
is the energy of the incorrectly identified $\pi$ meson. The $\ks$ background
is described by the function
\begin{equation}
B_{\ks}(\Phi) =
\exp\big[-\frac{(\sx-\sx(\ks))^2}{2\sigma^2}\big] P_\ks(\sy),
\end{equation}
where $P_\ks$ is a fourth-order polynomial and $\sigma$ is the resolution.

All the parameters in Eq.~\eqref{eq:bgpdf} are free except $\alpha_1$ and
the constant terms
of the polynomials $P_\mathrm{sm}$, $P_\varphi$ and $P_{\theta_{\jp}}$,
which are fixed at 1.
The $B\to\jp\ks$ events are present only in the left $\DE$ sideband.
This contribution is included in the fit of the sideband data
that is performed to determine the background shape but
excluded for the signal region.

\begin{figure}[t]
\begin{center}
\includegraphics[width=6cm]{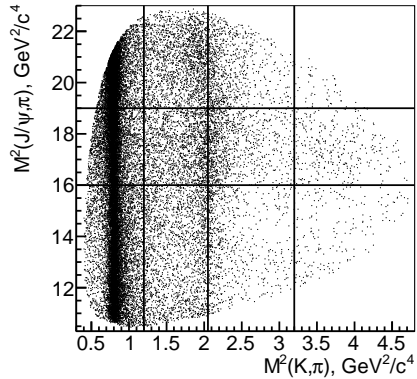}
\end{center}
\caption{Dalitz plot slices used to present results. Vertical
divisions are at $1.2\ \gevccsq$,
$(1.432\ \gevcc)^2\approx2.05\ \gevccsq$ and $3.2\ \gevccsq$
(the second division is chosen to be at the $K^*_2(1430)$ mass
since the interference of the $\kst$ resonances and the $\zp{4200}$
changes at this mass).
Horizontal divisions are at $16\ \gevccsq$ and $19\ \gevccsq$.}
\label{fig:slices}
\end{figure}

\begin{figure*}[t]
\begin{center}
\includegraphics[width=4.2cm,height=4.2cm]{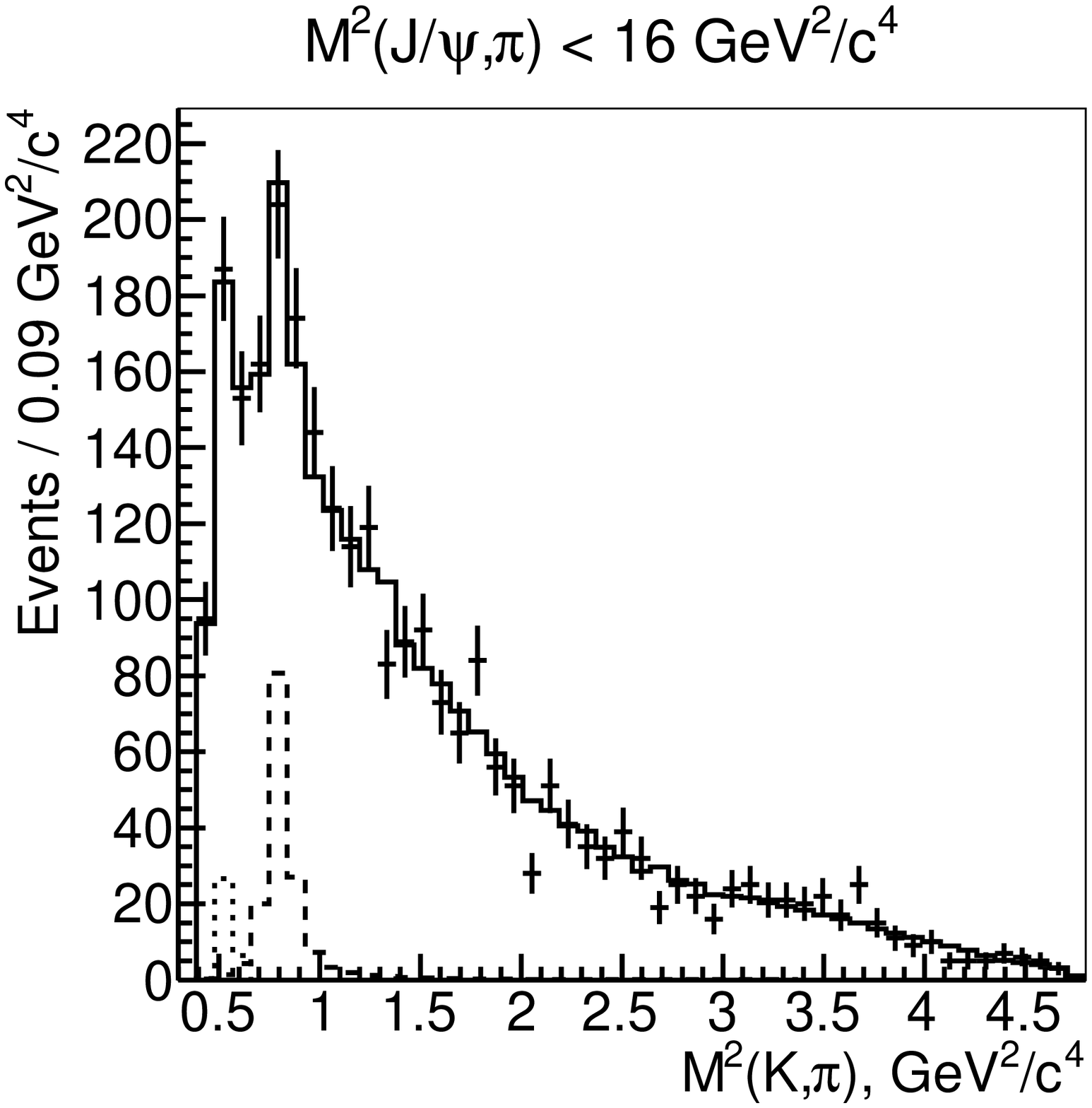}
\includegraphics[width=4.2cm,height=4.2cm]{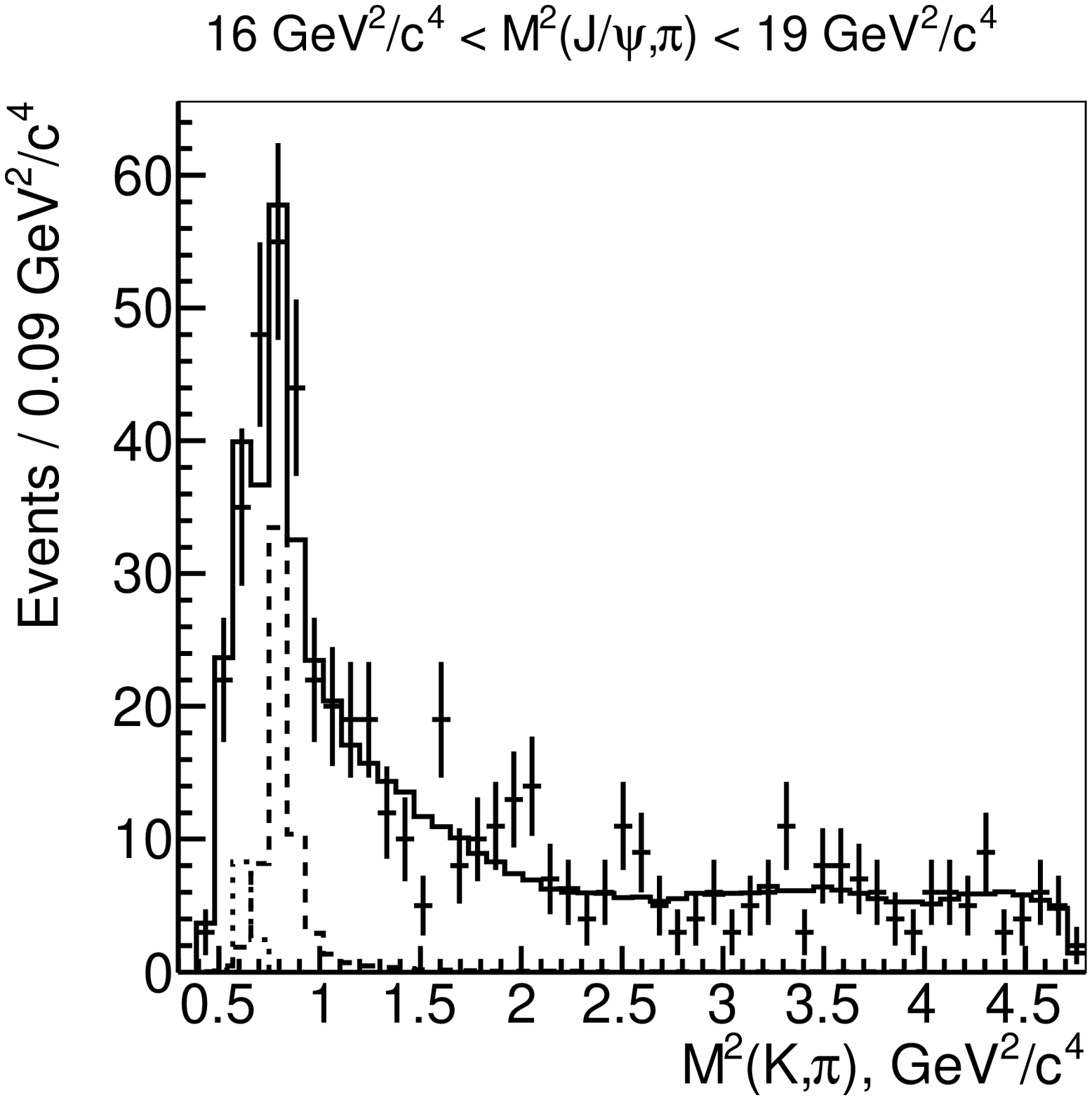}
\includegraphics[width=4.2cm,height=4.2cm]{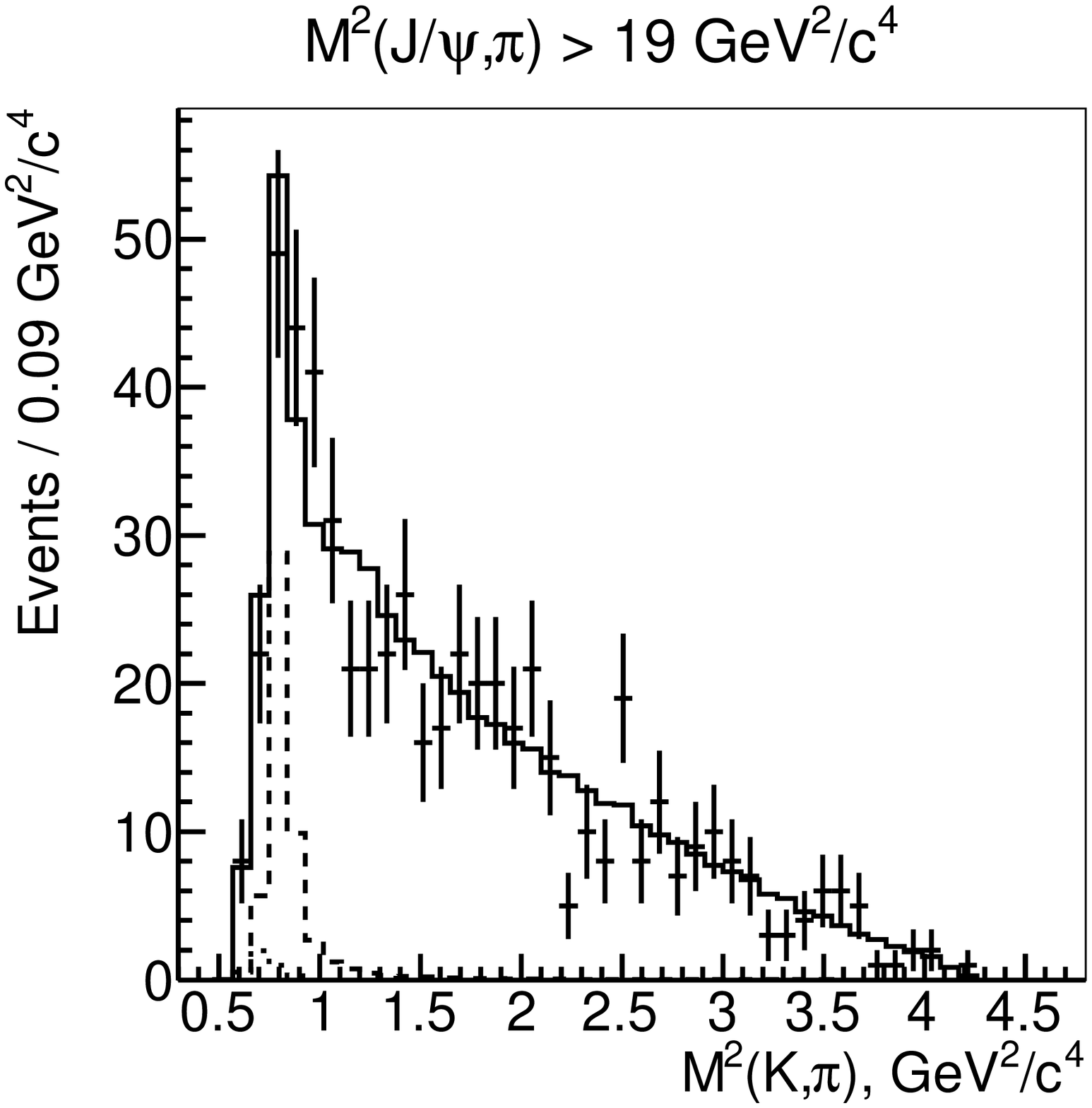}
\end{center}
\begin{center}
\includegraphics[width=4.2cm,height=4.2cm]{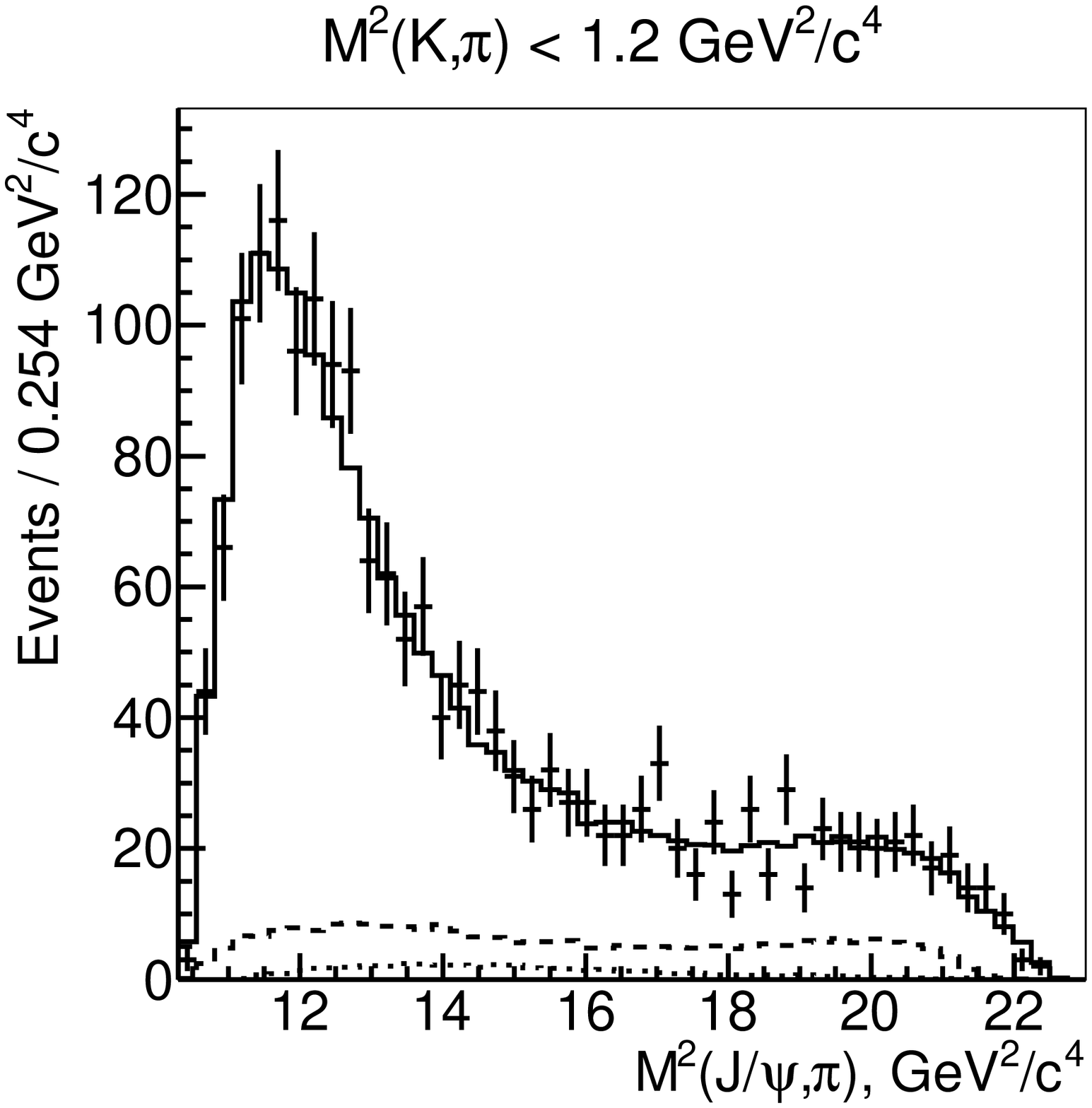}
\includegraphics[width=4.2cm,height=4.2cm]{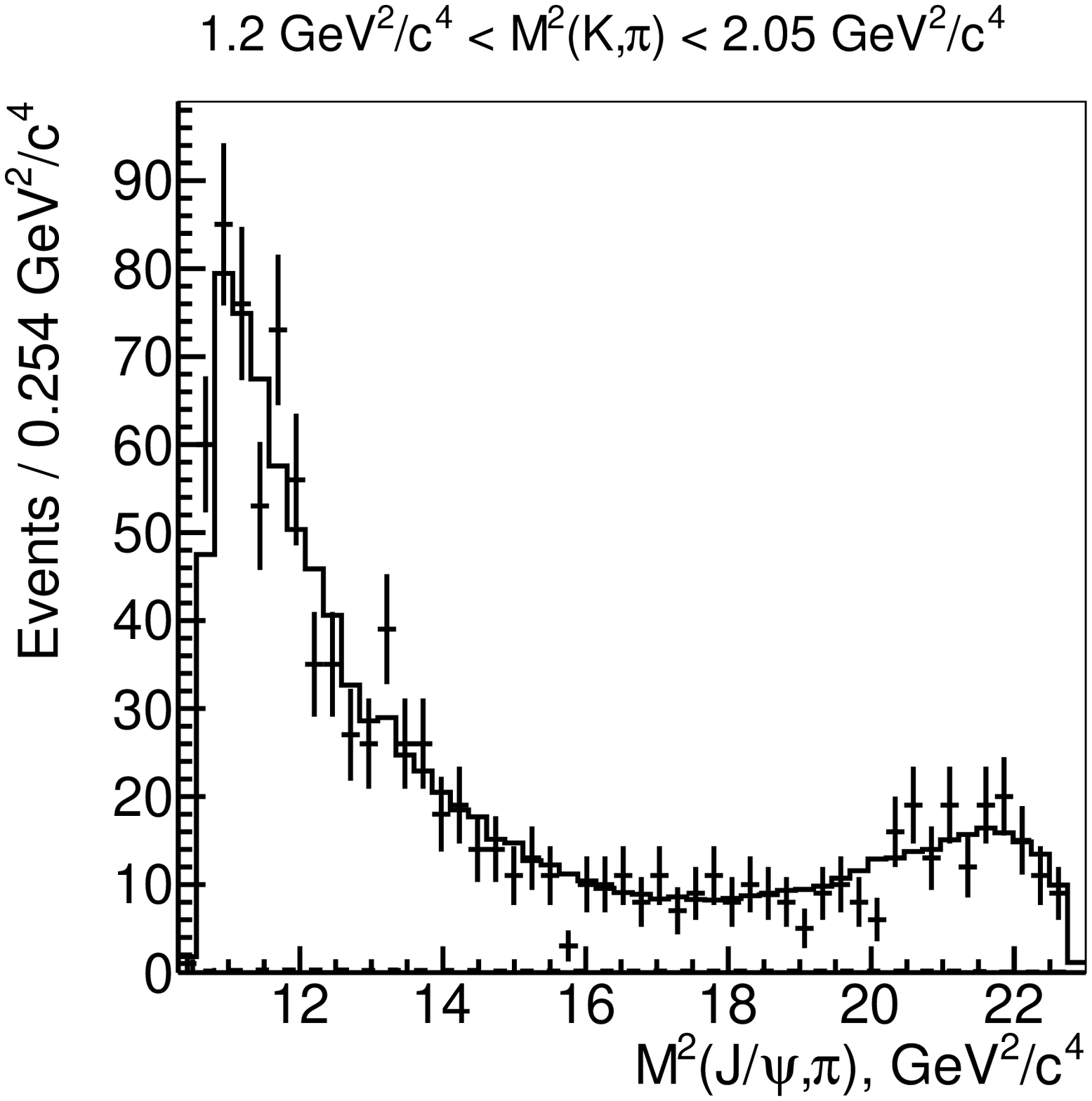}
\includegraphics[width=4.2cm,height=4.2cm]{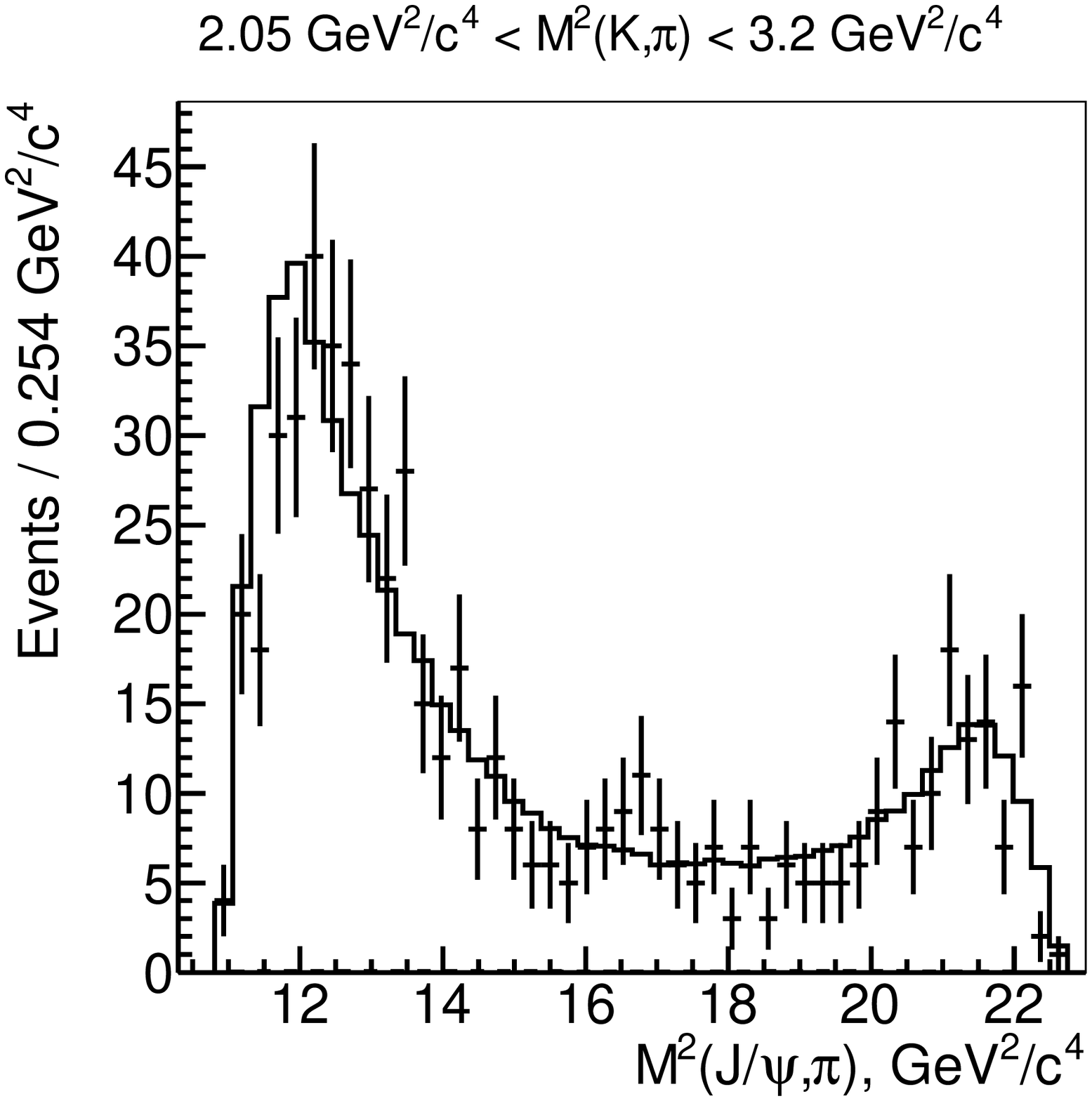}
\includegraphics[width=4.2cm,height=4.2cm]{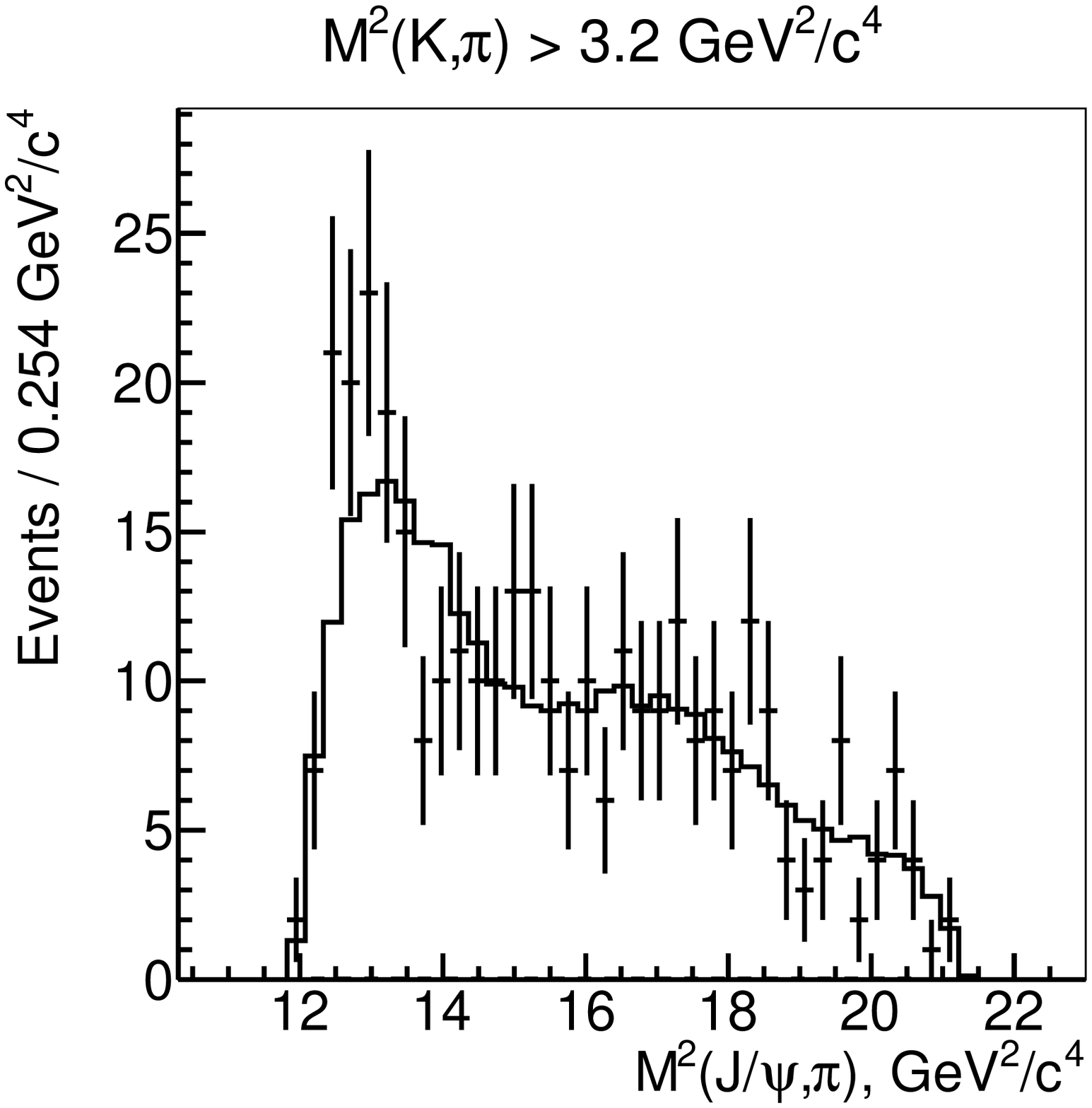}
\end{center}
\caption{Fit to the background events. The solid line is the fit result;
the dashed line is the $K^*(892)$ component;
the dotted line is the $K_S^0\to\pi^+\pi^-$ component.
The slices are defined in Fig.~\ref{fig:slices}.}
\label{fig:sbfitres}
\end{figure*}

We perform an unbinned maximum likelihood fit over
the four-dimensional space $\Phi$. The likelihood function is the same as
in Ref.~\cite{z4430jp}.
The masses and widths of all the $\kst$ resonances except $K^*_0(800)$
are fixed to their nominal values~\cite{PDG}.
The mass and width of the $K^*_0(800)$ are fixed to the fit results
in the default model without a $\z$
($M$ = $931\pm21\ \mevcc$, $\Gamma$ = $578\pm49\ \mev$);
the case of free mass and width is included in the systematic uncertainty.
The mass $M$ and the width $\Gamma$ of the $\zp{4430}$ are free parameters;
however, the known mass $M_0$ and width $\Gamma_0$ are used to
limit the floating mass and width by modifying $-2\ln L$:
\begin{equation}
-2\ln L \to -2\ln L + \frac{(M-M_0)^2}{\sigma_{M_0}^2}
+ \frac{(\Gamma-\Gamma_0)^2}{\sigma_{\Gamma_0}^2},
\label{eq:z4430masslim}
\end{equation}
where $\sigma_{M_0}$ and $\sigma_{\Gamma_0}$ are the uncertainties of $M_0$
and $\Gamma_0$, respectively.
The values of the $\zp{4430}$ mass and width are taken from Ref.~\cite{z4430jp}:
\begin{equation*}
M_0 = 4485^{+36}_{-25}\ \mevcc,\ \Gamma_0 = 200^{+49}_{-58}\ \mev.
\end{equation*}
Other details of the fitting procedure are the same as in Ref.~\cite{z4430jp}.


\section{Results}

\subsection{Fit results}

The background shape is determined from an unbinned maximum likelihood
fit to the events in the $\DE$ sidebands.
To present the fit results, the Dalitz plot is divided into the slices
shown in Fig.~\ref{fig:slices}. The results of the fit to the background
events are shown in Fig.~\ref{fig:sbfitres}.


A search for a $\z$ with arbitrary mass and width is performed. The considered
spin-parity hypotheses are $J^P=0^-$, $1^-$, $1^+$, $2^-$ and $2^+$.
The $0^+$ combination is forbidden by
parity conservation in $\z\to\jp\pip$ decays.
The fit results for the $\z$ mass, width and significance
in the default model are shown in
Table~\ref{tab:z_jp}.
The Wilks significance of the $\z$ with $J^P=1^+$ is
$8.2\sigma$; its global significance is $7.9\sigma$. The
significance calculation method is described in Appendix~\ref{sec:significance}.
The global significance with the systematic uncertainty is
$6.2\sigma$ (the calculation is described further in this section).
Thus a new state, referred to in the following as the $\zp{4200}$,
is observed.
The preferred spin-parity hypothesis is $1^+$.
We also see a signal for $\zp{4430}\to\jp\pip$ with a Wilks significance
of $5.1\sigma$ in the default model;
the global significance is found to be the same.
The significance with the systematic uncertainty is $4.0\sigma$.
Thus we find evidence for a new decay channel of the $\zp{4430}$.

To test the goodness of the fit, we bin the Dalitz distribution with the
requirement that the number of events in each bin satisfy $n_i>25$.
We then calculate
the $\chi^2$ value as $\sum_i (n_i-s_i)^2/s_i$, where $s_i$ is the integral of
the fitting function (the result of unbinned fit)
over bin $i$. Since the fit is a maximum likelihood fit,
we obtain the effective number of degrees of freedom by
generating
MC pseudoexperiments in accordance with the result of the fit; then,
the distribution of the $\chi^2$ value in
the pseudoexperiments is fitted to the $\chi^2$ distribution
with variable number of degrees of freedom.
The confidence level of the fit with the $\zp{4200}$
(for the $1^+$ hypothesis) is 13\%; the confidence level
of the fit without the $\zp{4200}$ is 1.8\%.
We also calculate the confidence level using four-dimensional binning
(three bins in $|\cos\theta_{\jp}|$, three bins in $\varphi$ and similar
adaptive binning for the Dalitz plot variables);
the resulting confidence level is larger.
The amplitude absolute values and phases in the default model are listed in
Table~\ref{tab:defamp}.
The significances of the $K^*$ resonances are shown in Table~\ref{tab:ffrac}.

\begin{table*}
\caption{
Fit results in the default model. Errors are statistical only.
}
\begin{tabular}{c|c|c|c|c|c}
\hline\hline
$J^P$ & $0^-$ & $1^-$ & $1^+$ & $2^-$ & $2^+$ \\
\hline
Mass, $\mevcc$ & $4318\pm48$ & $4315\pm40$ & $4196^{+31}_{-29}$ & $4209\pm14$ & $4203\pm24$ \\
Width, $\mev$ & $720\pm254$ & $220\pm80$ & $370\pm70$ & $64\pm18$ & $121\pm53$ \\
Significance 
(Wilks) & $3.9\sigma$ & $2.3\sigma$ & $8.2\sigma$ & $3.9\sigma$ & $1.9\sigma$ \\
\hline\hline
\end{tabular}
\label{tab:z_jp}
\end{table*}

\begin{figure*}
\includegraphics[width=6cm]{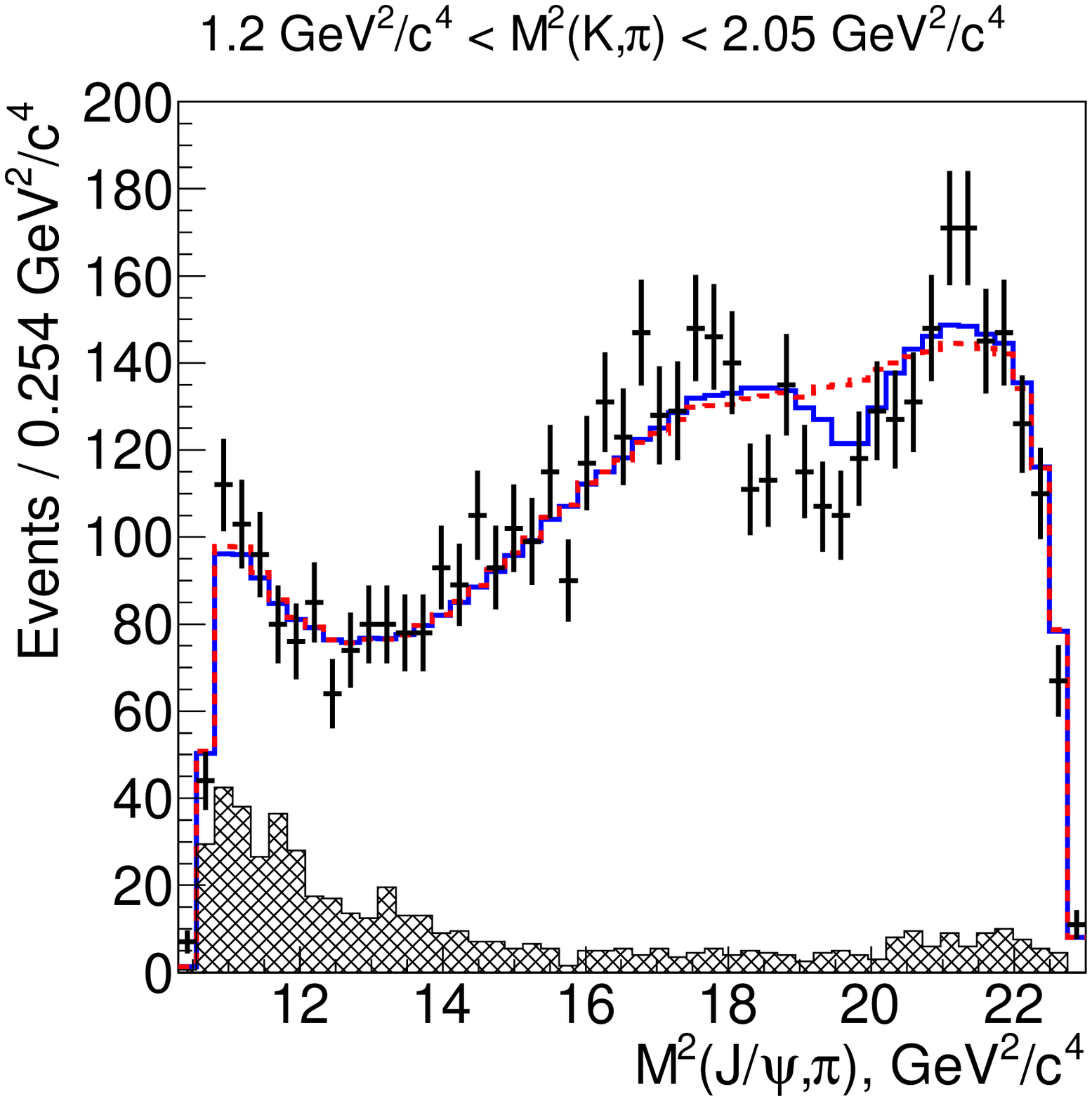}
\includegraphics[width=6cm]{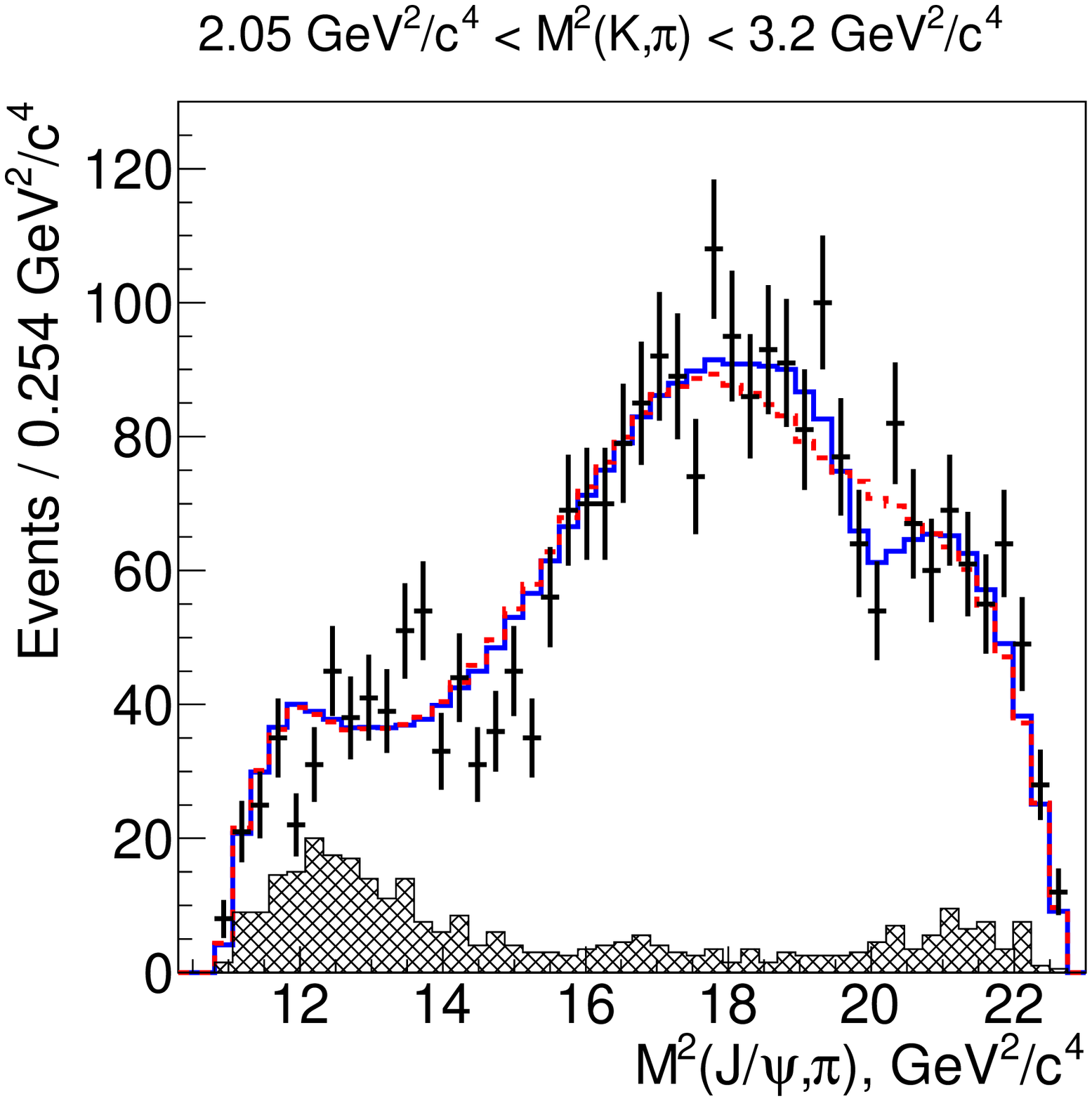}
\caption{The fit results with (solid line) and without (dashed line)
the $\zp{4430}$ (the $\zp{4200}$ is not included
in the model) for the second and third vertical slices that are defined in
Fig.~\ref{fig:slices}.}
\label{fig:z4430}
\end{figure*}

\begin{figure*}[ht]
\begin{center}
\includegraphics[width=4.2cm,height=4.2cm]{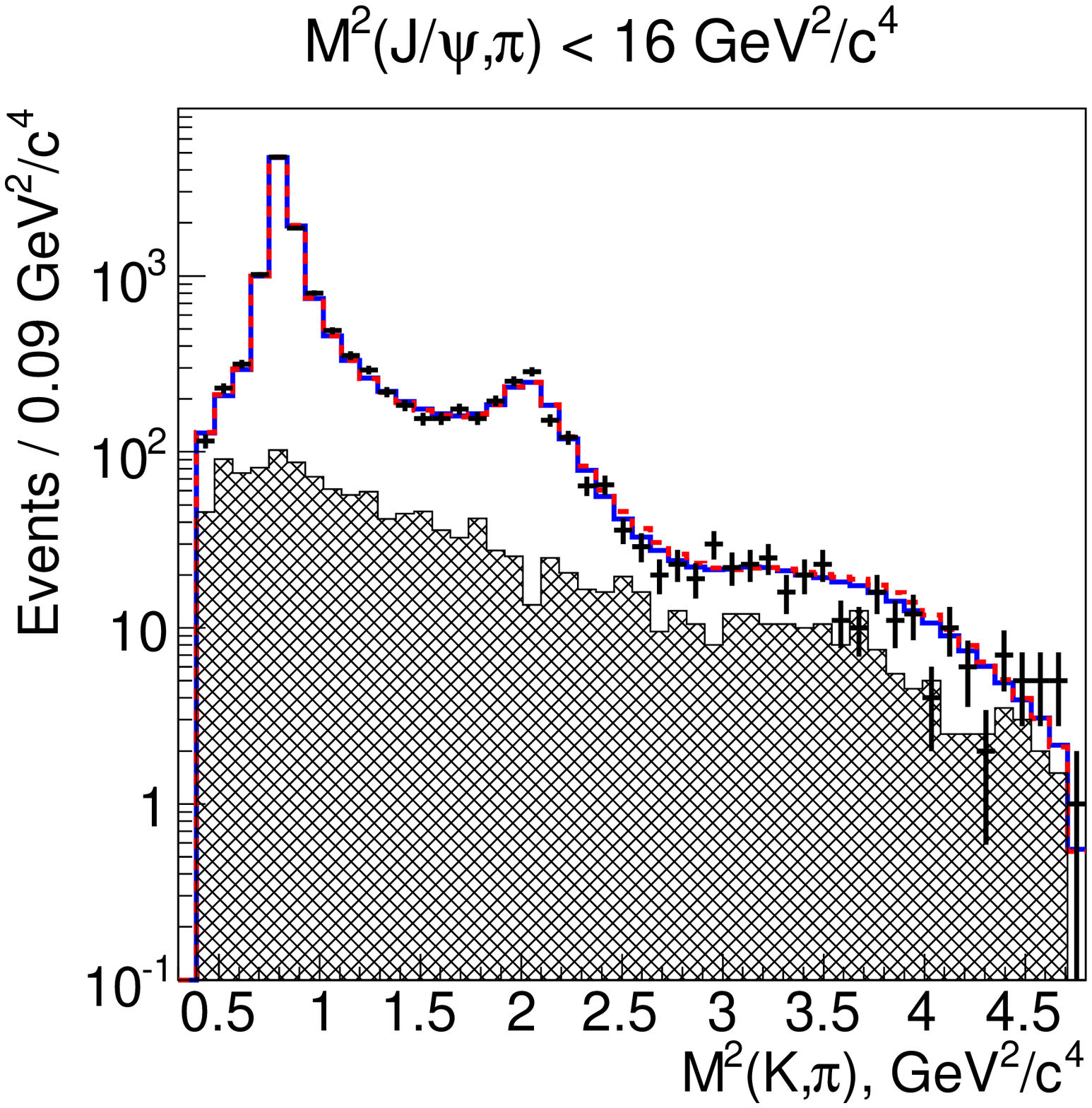}
\includegraphics[width=4.2cm,height=4.2cm]{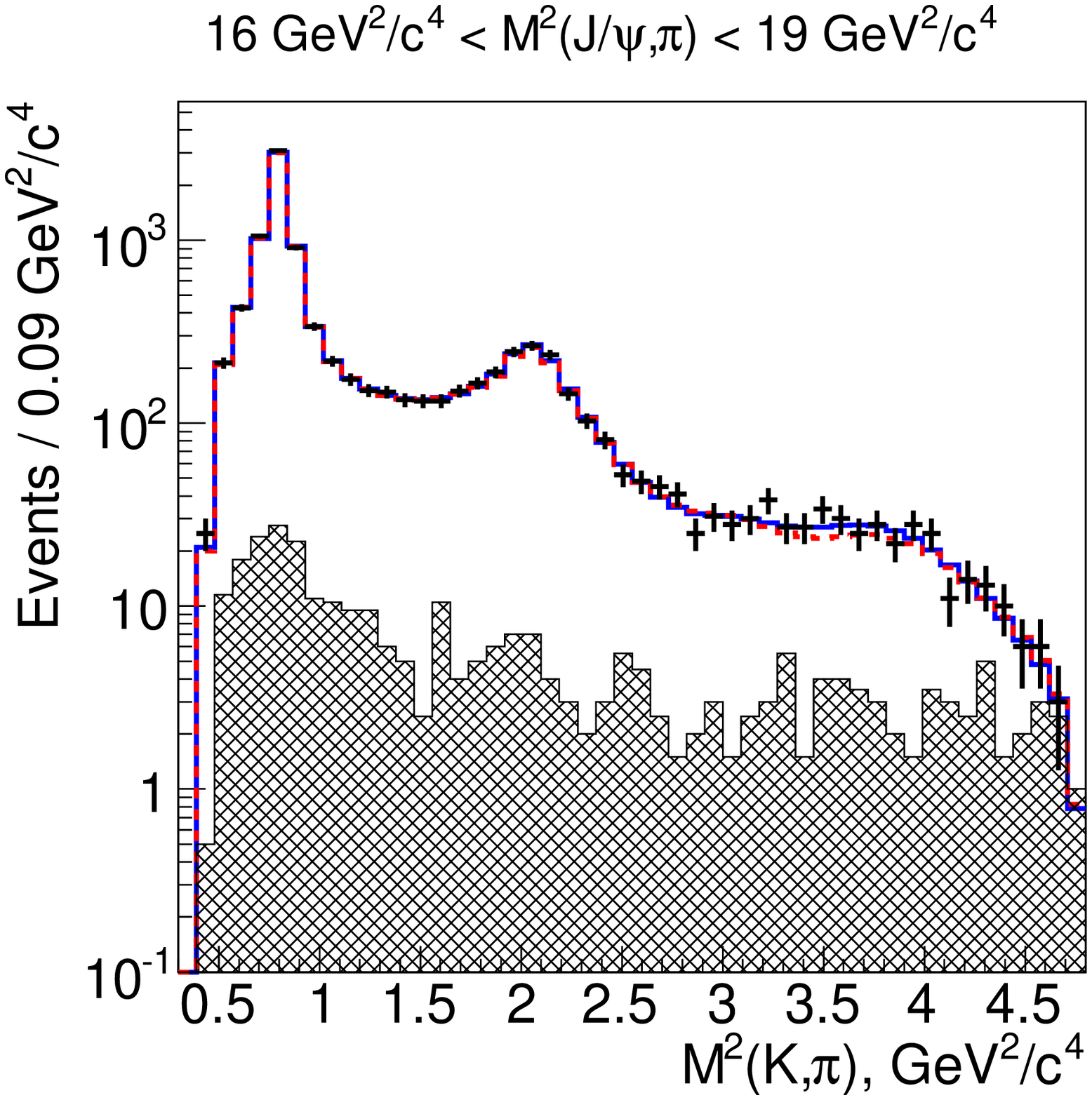}
\includegraphics[width=4.2cm,height=4.2cm]{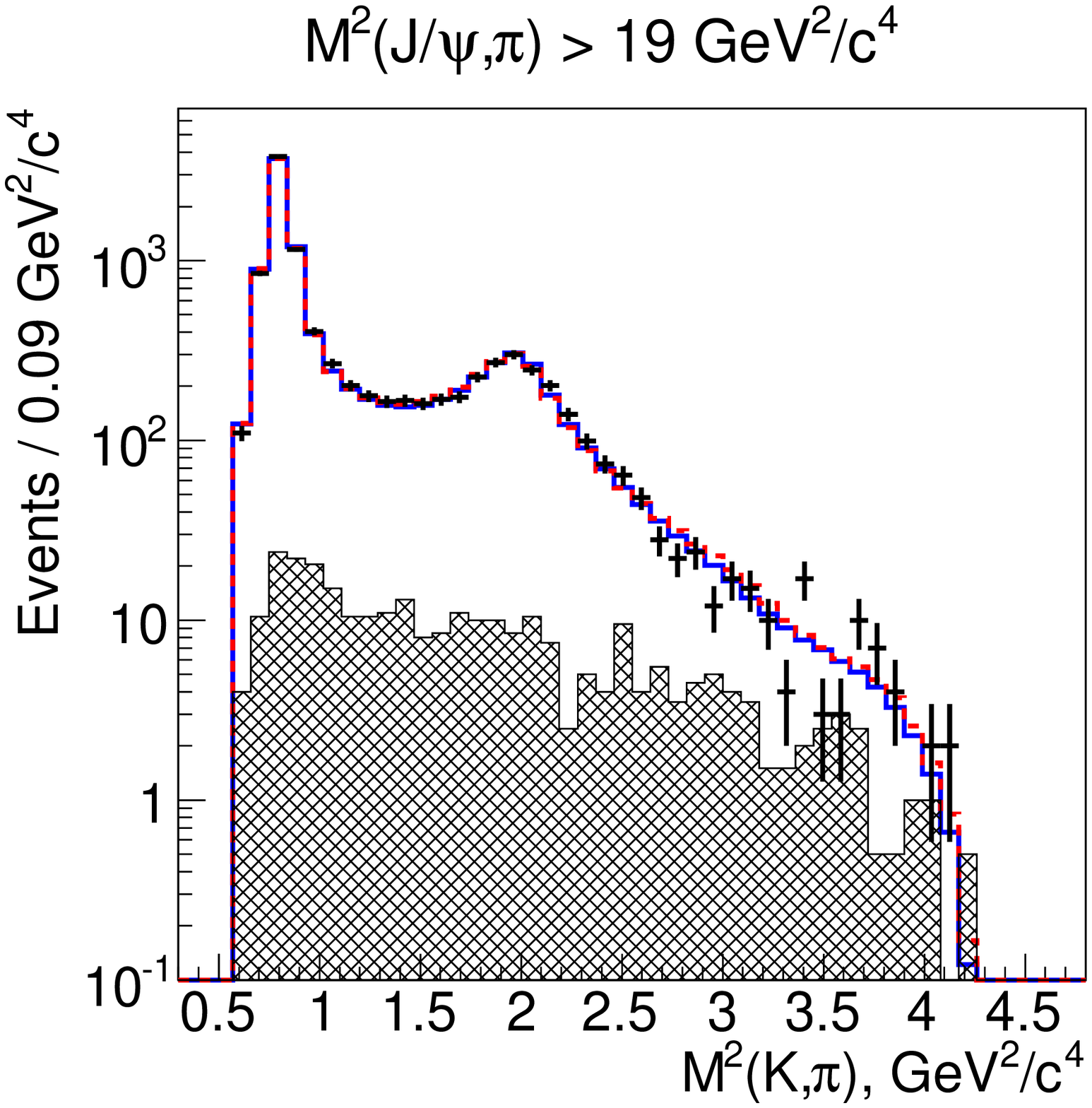}
\end{center}
\begin{center}
\includegraphics[width=4.2cm,height=4.2cm]{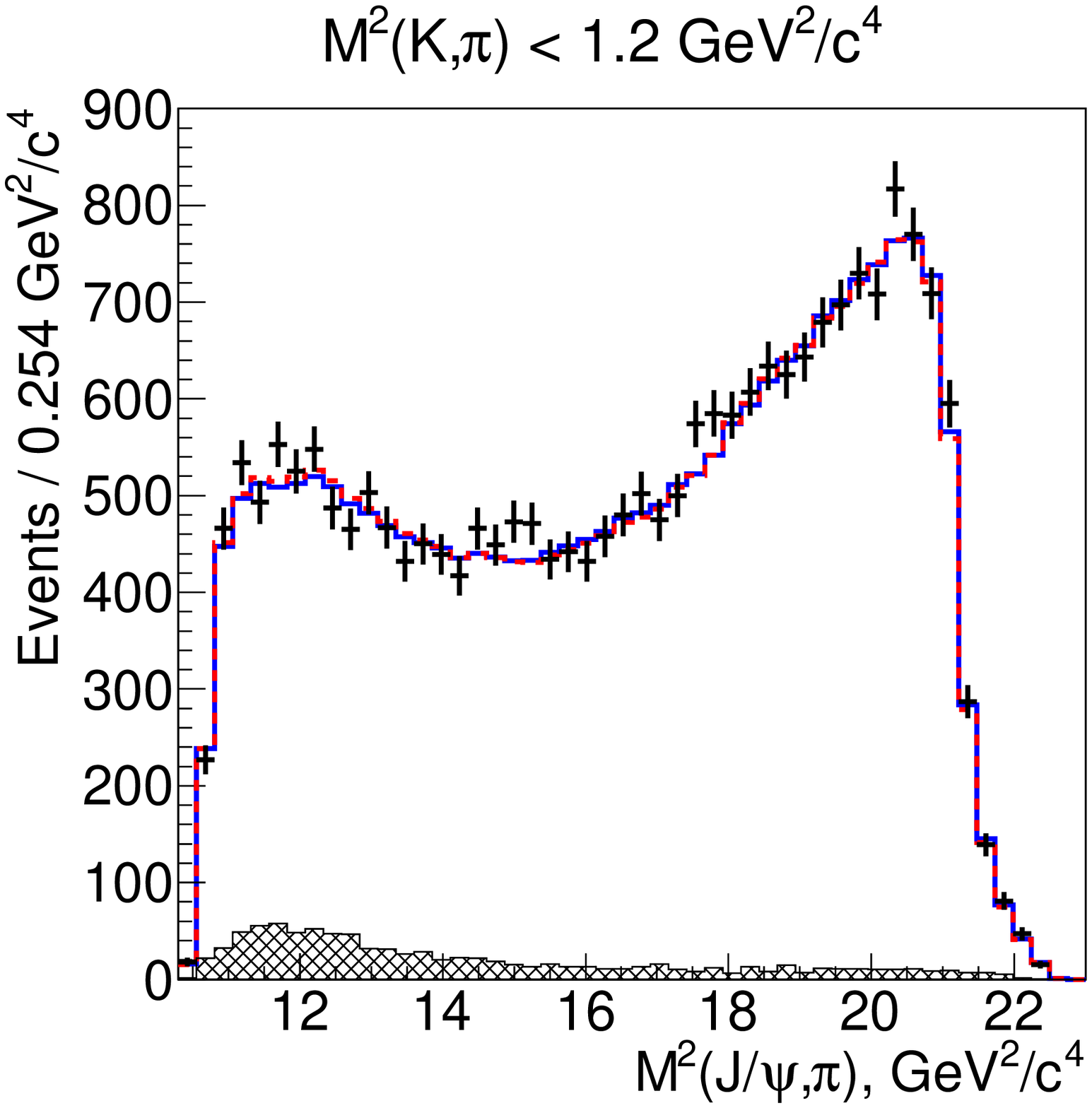}
\includegraphics[width=4.2cm,height=4.2cm]{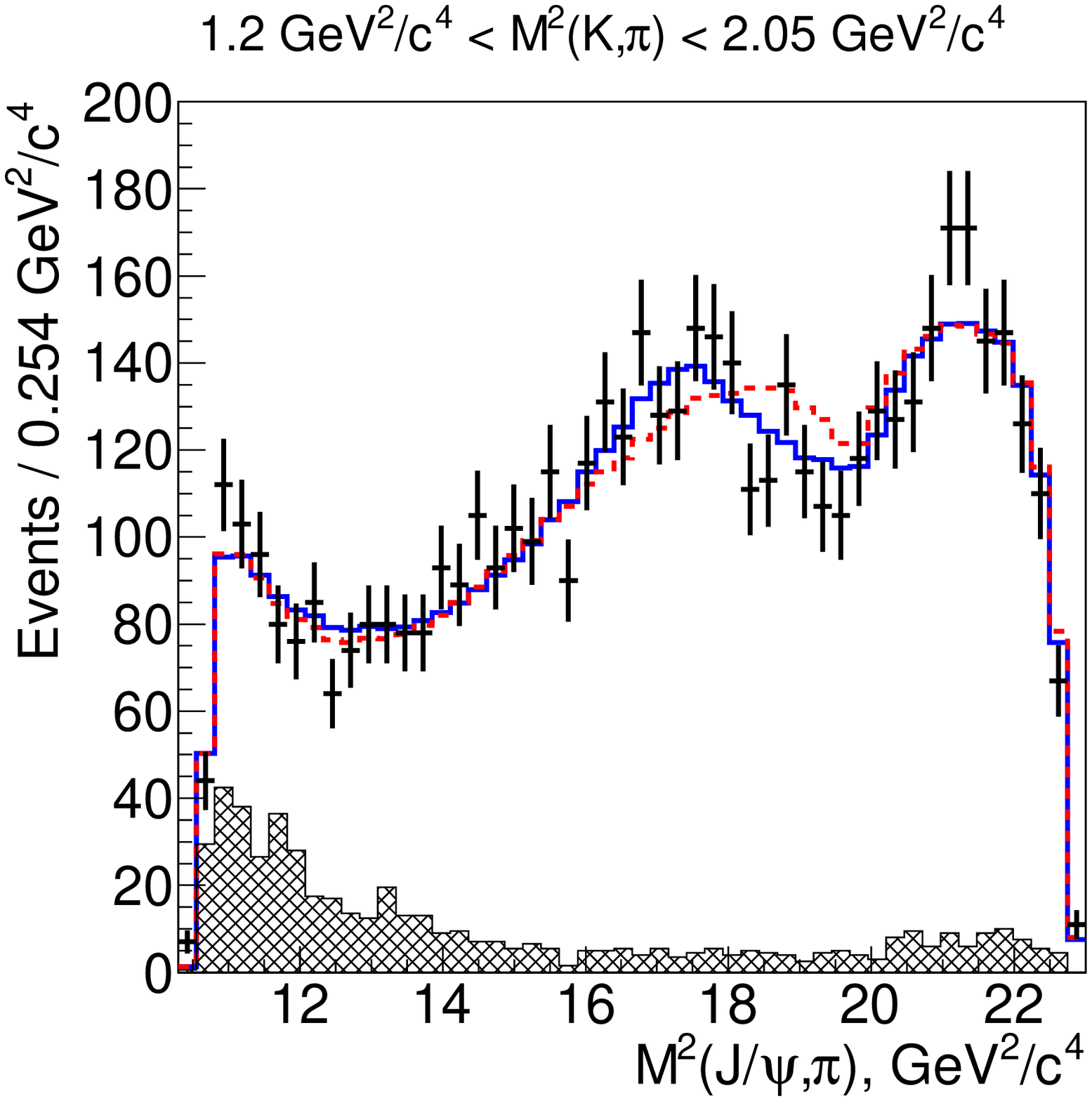}
\includegraphics[width=4.2cm,height=4.2cm]{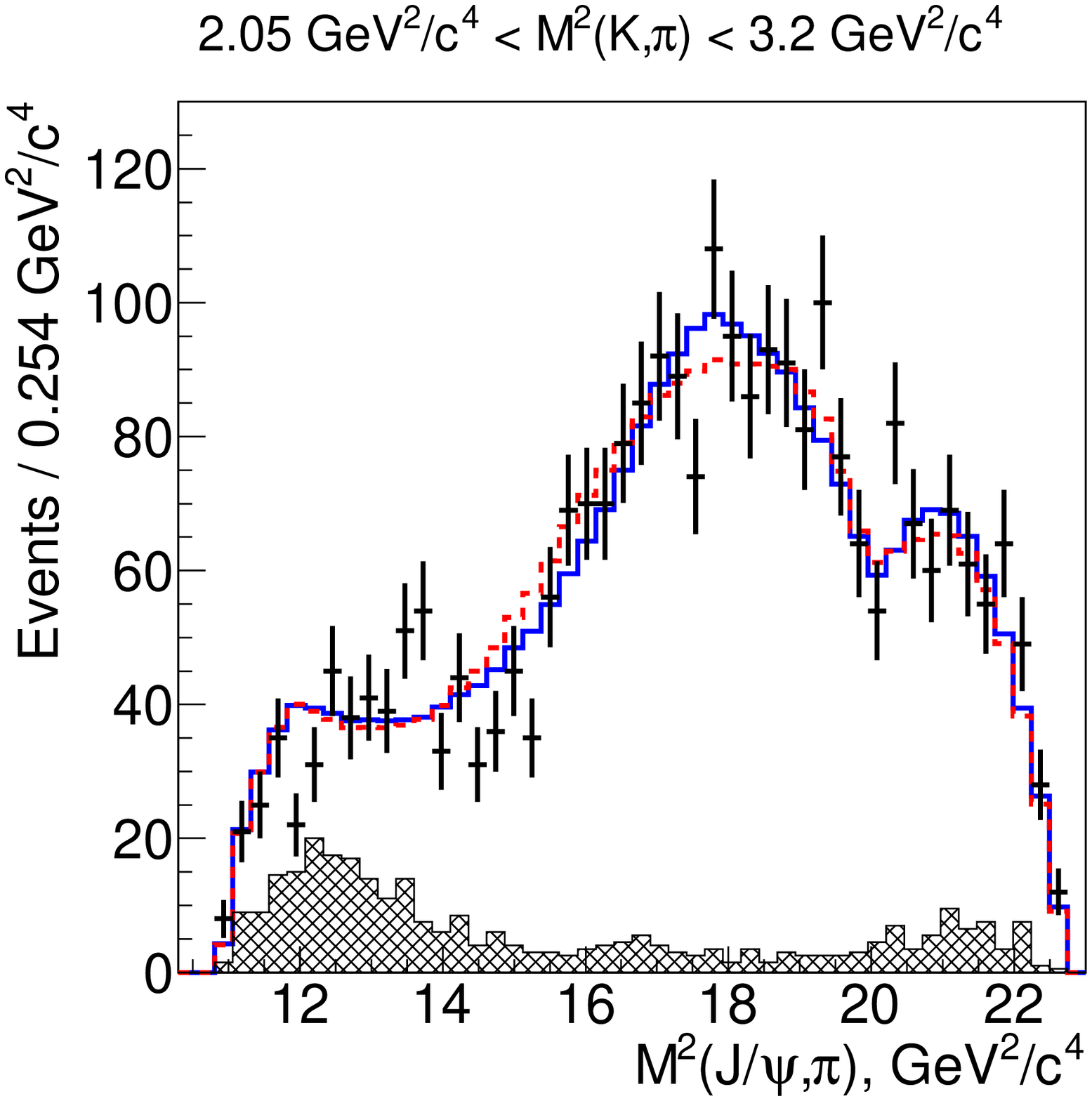}
\includegraphics[width=4.2cm,height=4.2cm]{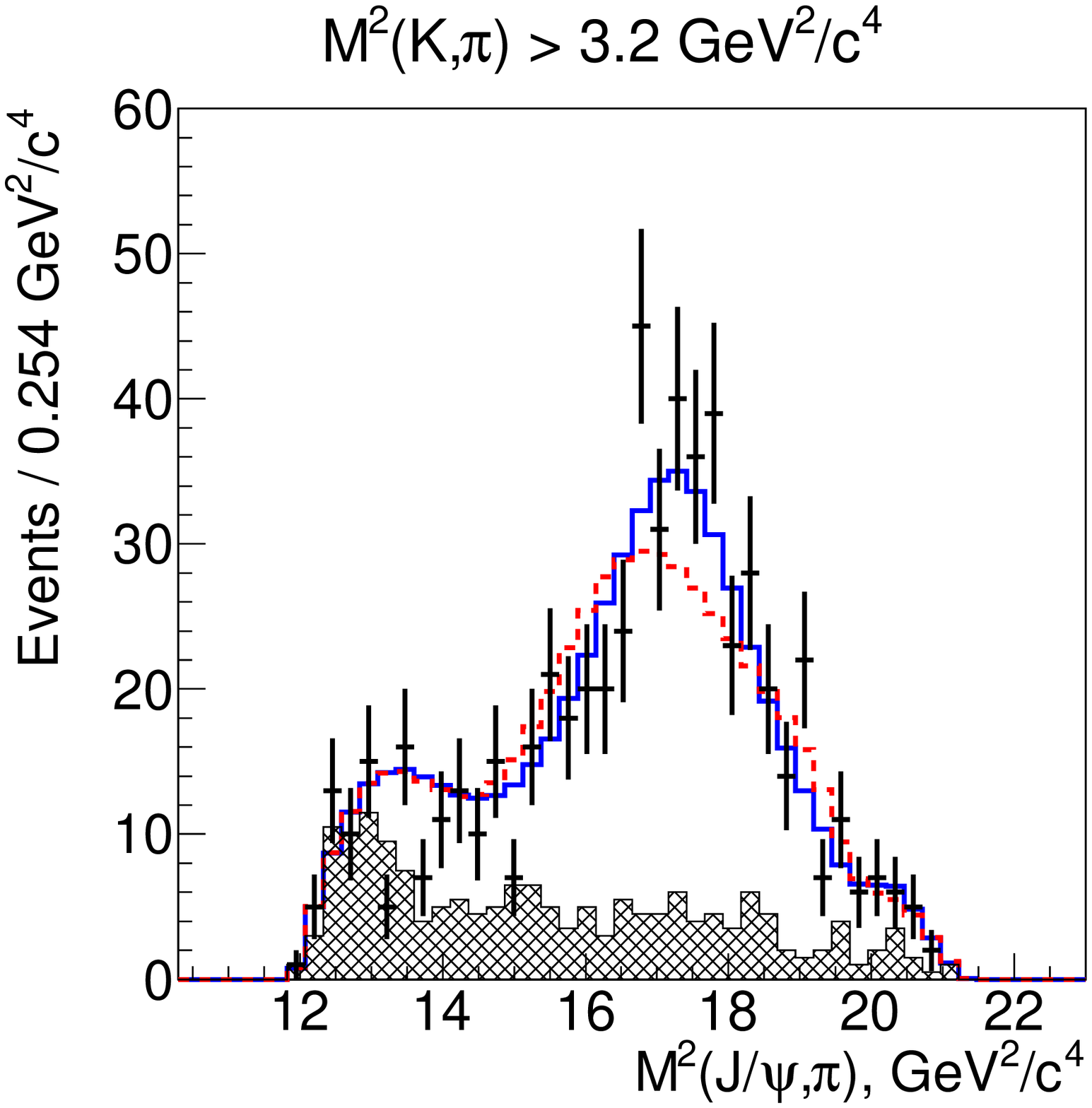}
\end{center}
\caption{The fit results with (solid line) and without (dashed line)
the $\zp{4200}$ ($J^P=1^+$) in the default model.
The points with error bars are data;
the hatched histograms are the $\jp$ sidebands. The slices are defined in
Fig.~\ref{fig:slices}.}
\label{fig:fitresdef}
\end{figure*}

\begin{figure*}[ht]
\begin{center}
\includegraphics[width=5.5cm]{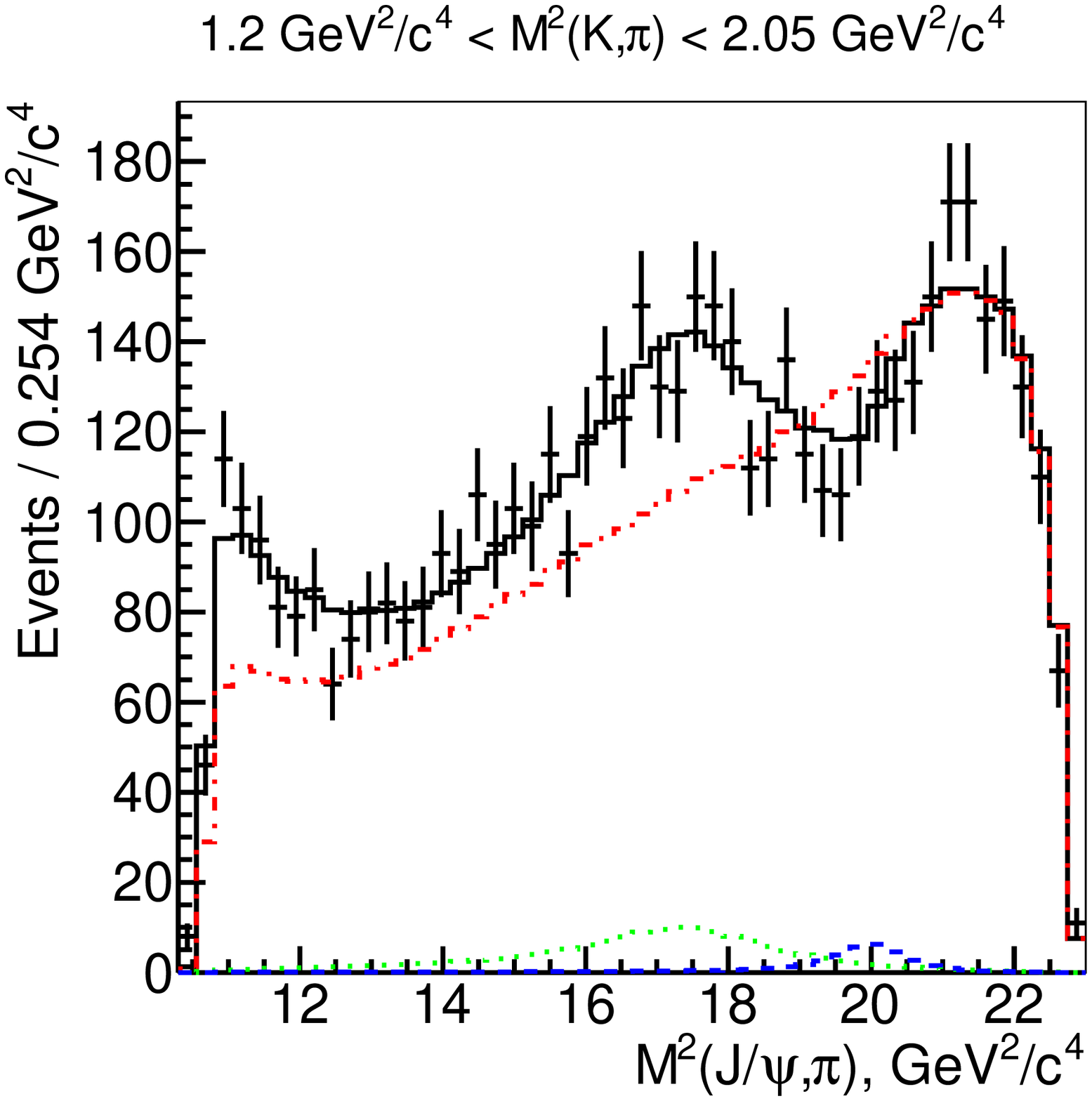}
\includegraphics[width=5.5cm]{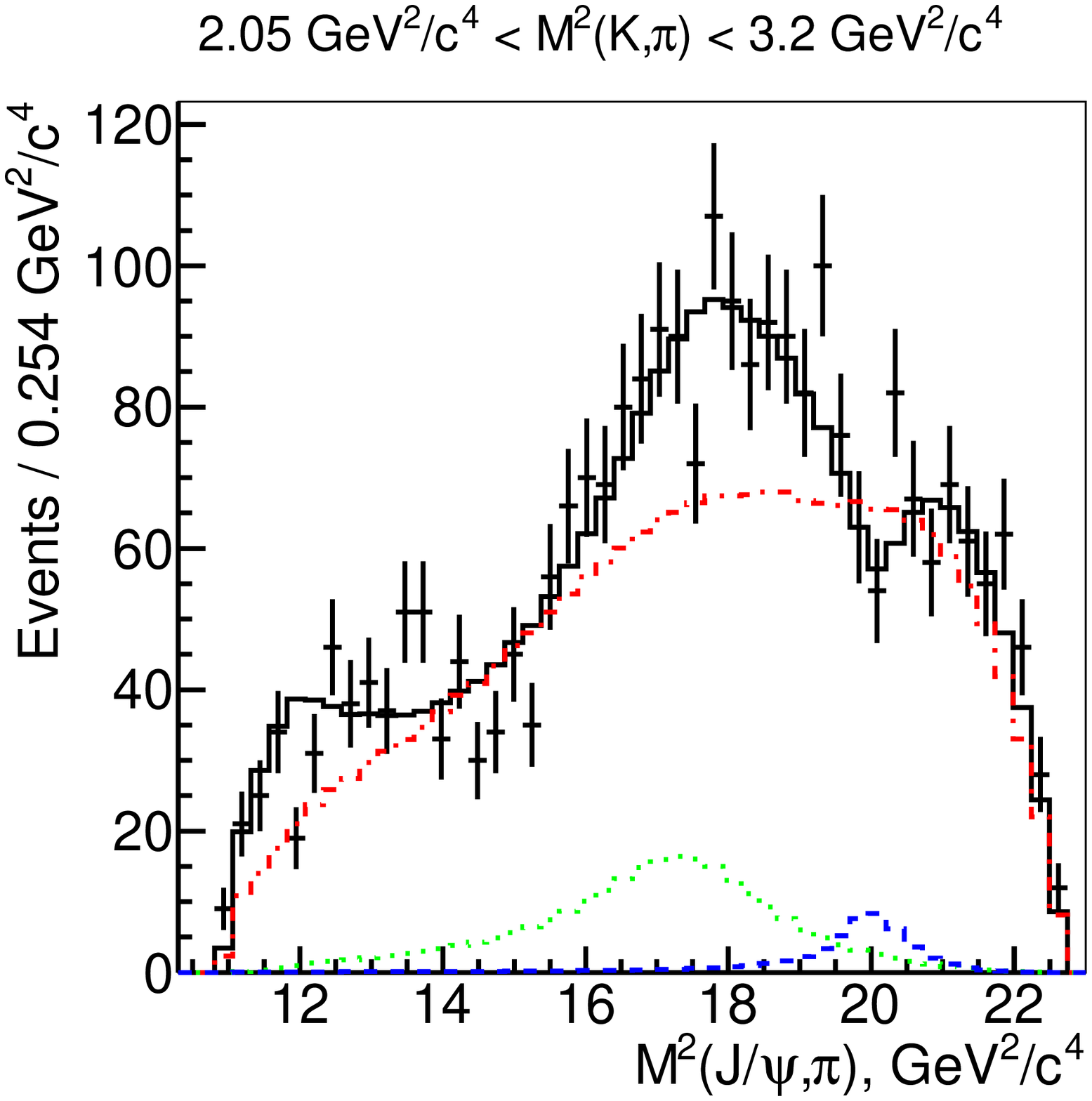}
\includegraphics[width=5.5cm]{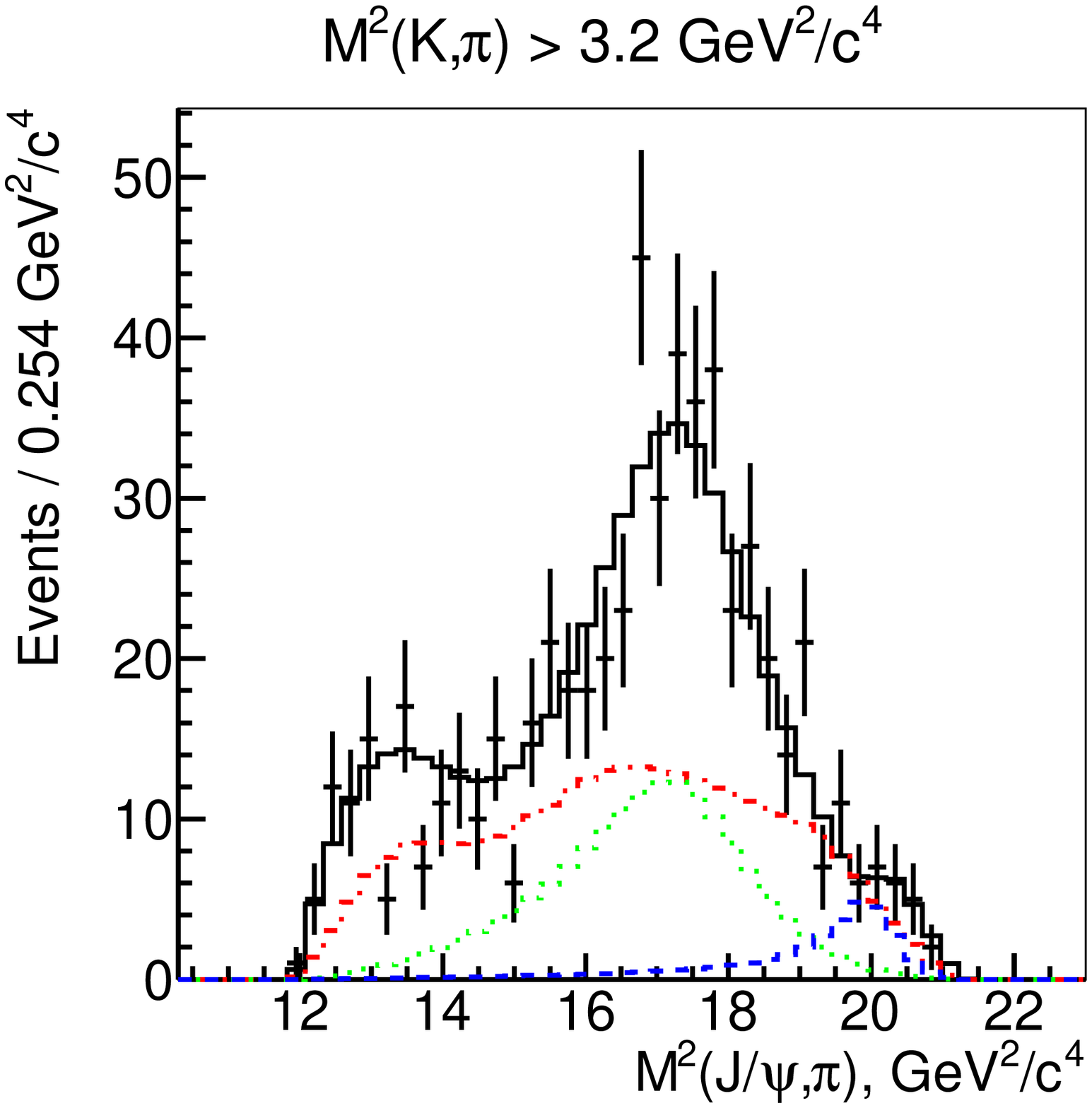}
\end{center}
\caption{The fit results with the $\zp{4200}$ ($J^P=1^+$) in the default model.
The points with error bars are data;
the solid histograms are fit results, the dashed histograms are the $\zp{4430}$
contributions, the dotted histograms are the $\zp{4200}$ contributions
and the dash-dotted histograms are contributions of all $\kst$ resonances.
The slices are defined in Fig.~\ref{fig:slices}.}
\label{fig:fitresdef_contrib}
\end{figure*}

\begin{figure}[ht]
\begin{center}
\includegraphics[width=6cm]{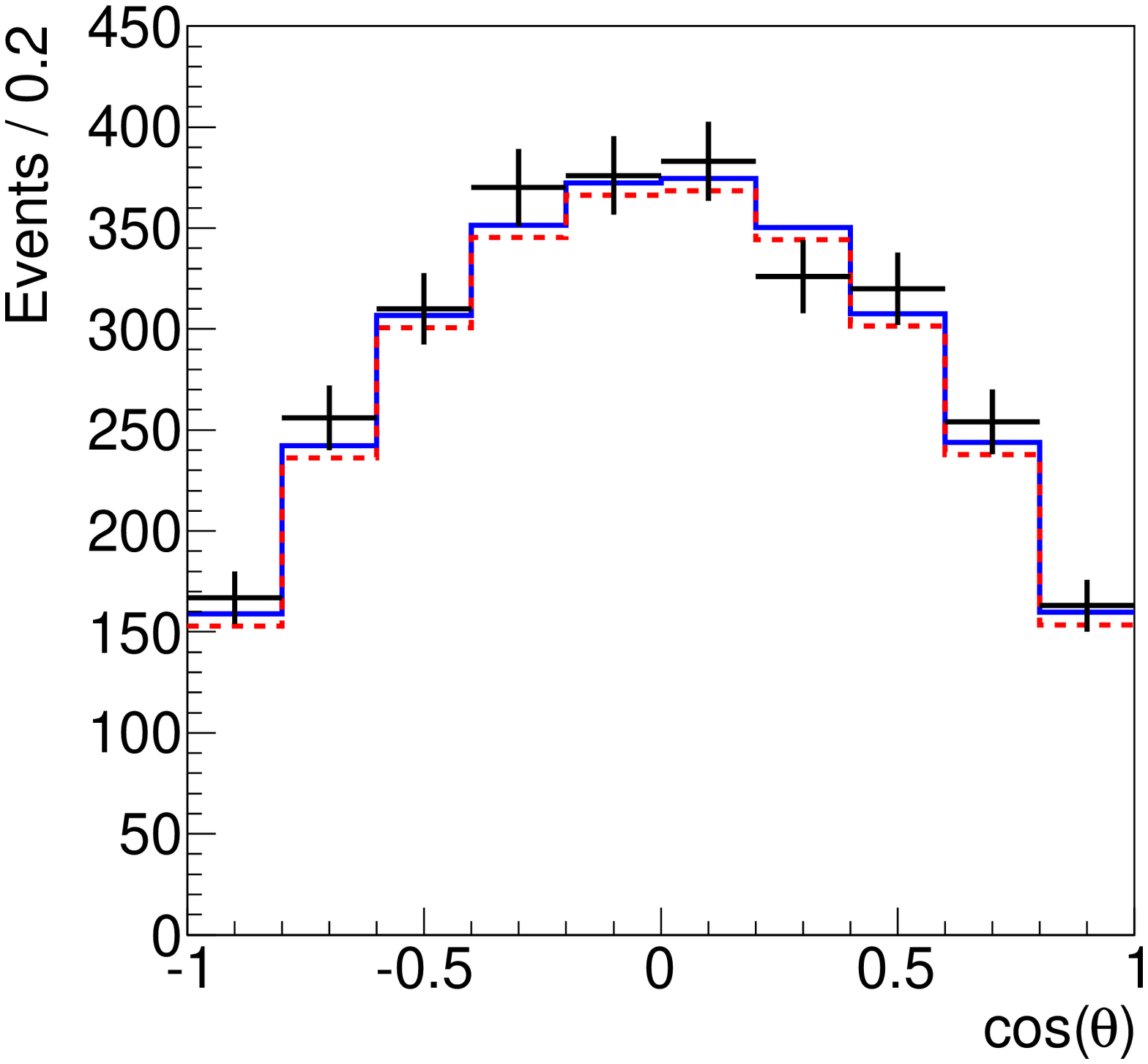}
\includegraphics[width=6cm]{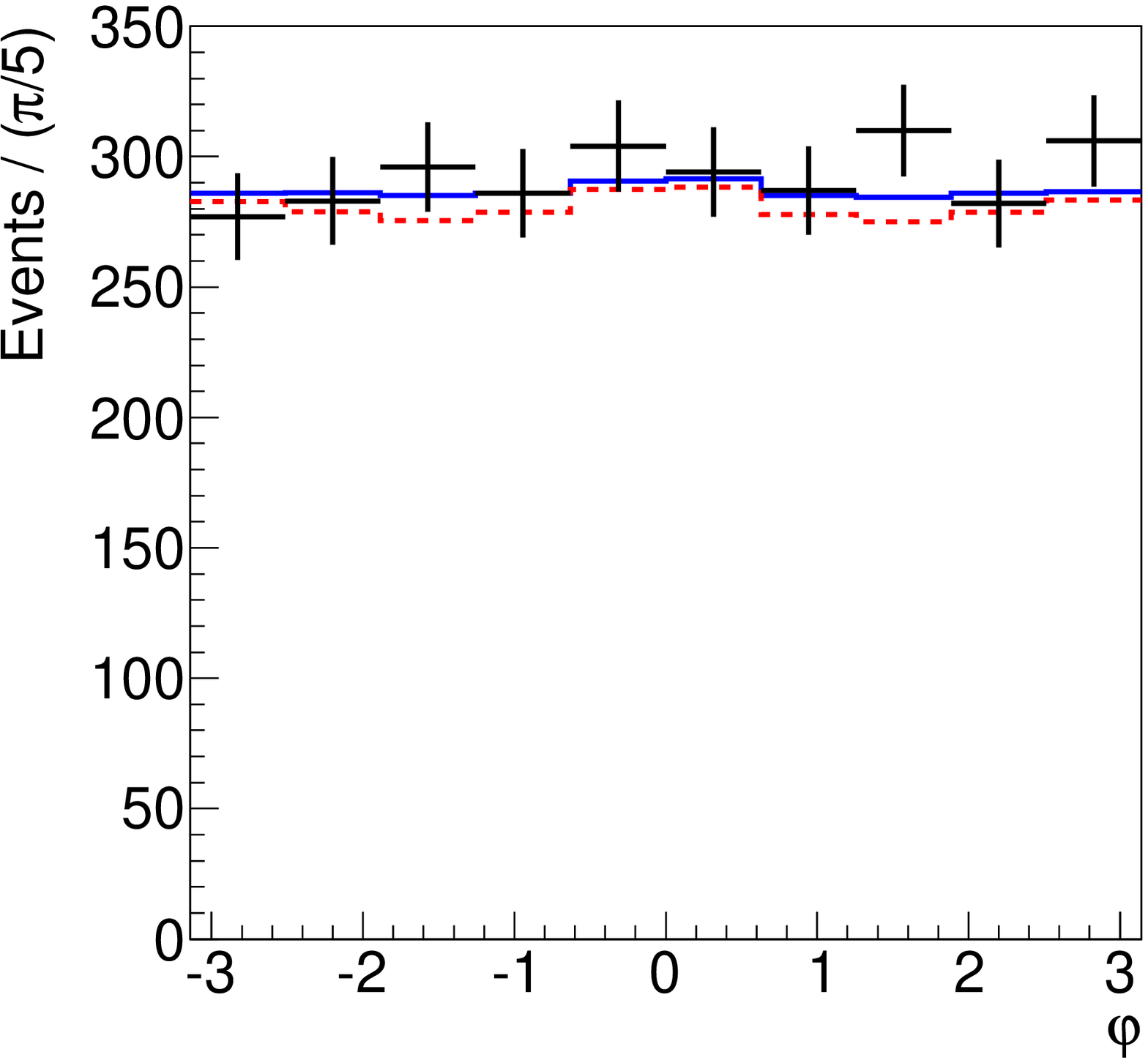}
\end{center}
\caption{Projections of the fit results with (solid line) and without
(dashed line) the $\zp{4200}$ ($J^P=1^+$) onto the angular variables in
the default model for the region defined by $\sx>1.2\,\gevccsq$,
$16\,\gevccsq<\sy<19\,\gevccsq$.
Points with error bars are data.}
\label{fig:fitresang}
\end{figure}

\begin{figure*}
\includegraphics[width=6cm]{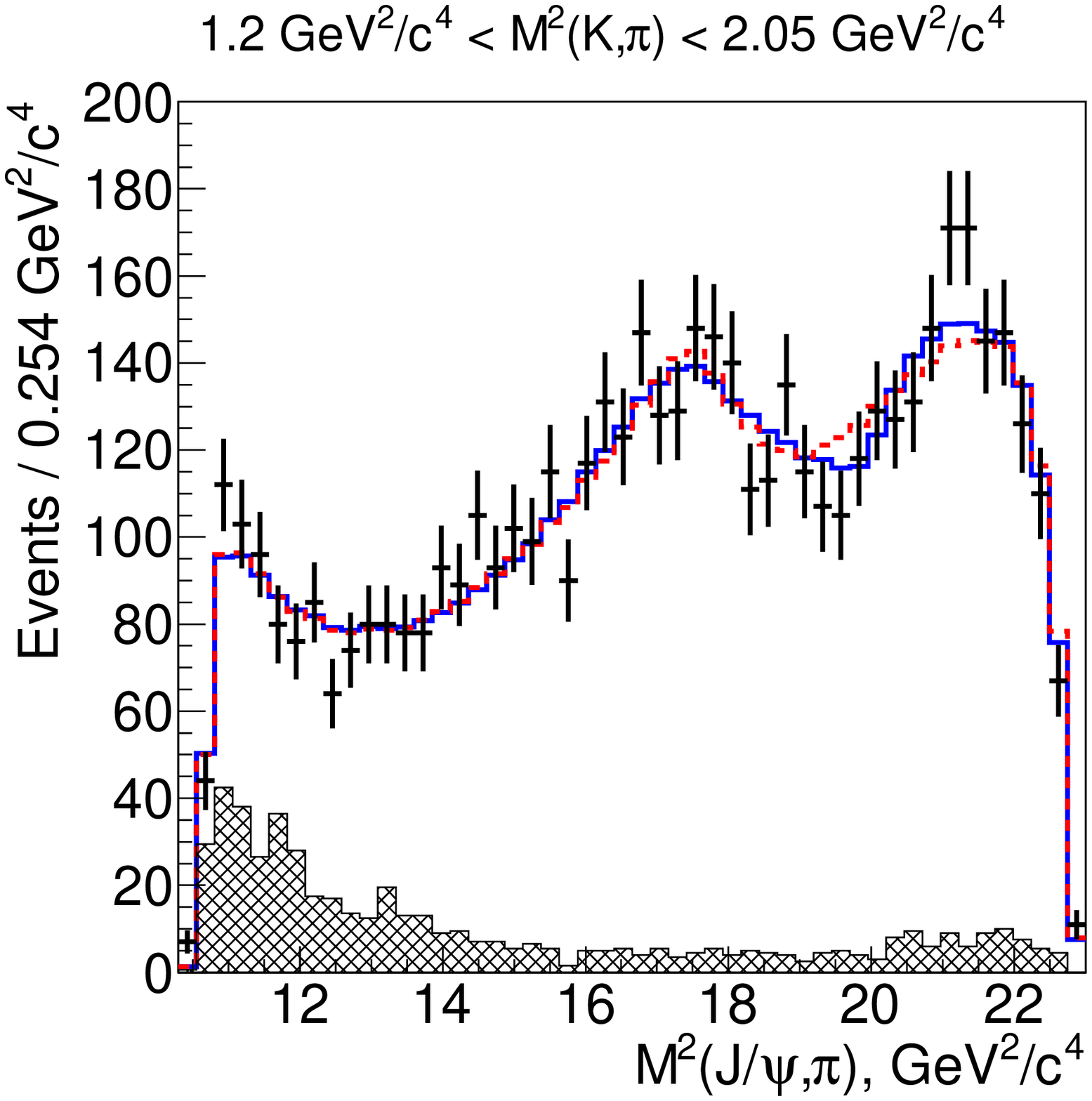}
\includegraphics[width=6cm]{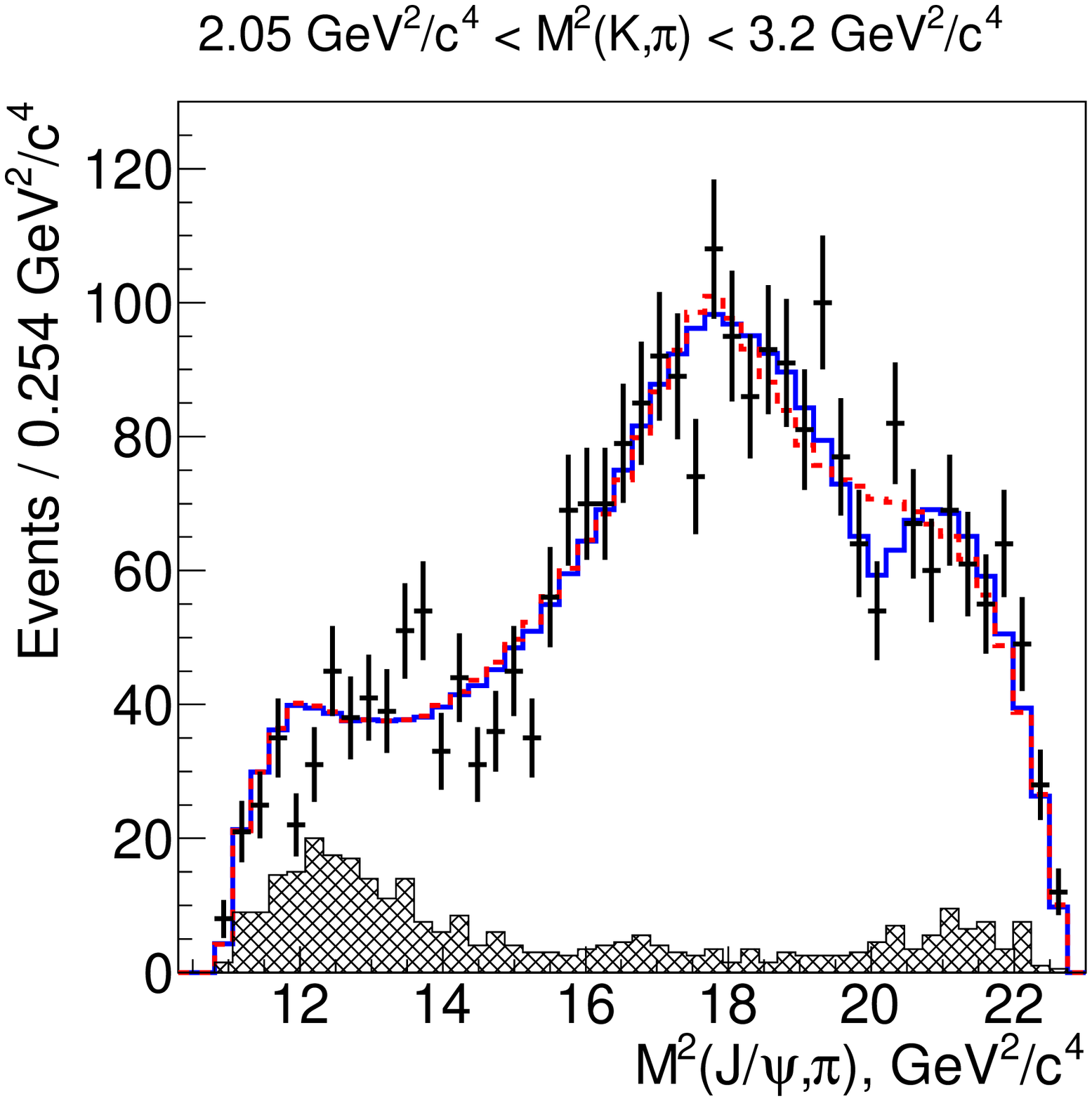}
\caption{The fit results with (solid line) and without (dashed line)
the $\zp{4430}$ (the $\zp{4200}$ is included
in the model) for the second and third vertical slices that are defined in
Fig.~\ref{fig:slices}.}
\label{fig:z4430_2}
\end{figure*}

Since the $\zp{4430}$ is a known resonance,
before showing the fit results with and without the $\zp{4200}$,
we present a comparison of the fit results with and without the
$\zp{4430}$ with the $\zp{4200}$ not included in the model,
as shown in Fig.~\ref{fig:z4430}. There is no peak in the $\zp{4430}$
region; instead, effects of destructive interference are seen.
Projections of the fit results
onto the $M^2_{K \pi}$ and $M^2_{\jp \pi}$ axes
for the model with the $\zp{4200}$ ($J^P=1^+$) and
the model without the $\zp{4200}$ are shown in Fig.~\ref{fig:fitresdef}.
The two peaks evident in the projections onto the $M^2_{K \pi}$ axis are due to
the $\kst(892)$ and $K^*_2(1430)$ resonances. The new resonance
$\zp{4200}$ is seen as a wide peak near the center of the projections
onto the $M^2_{\jp \pi}$ axis. Projections of the $K^*$, $\zp{4200}$ and
$\zp{4430}$ contributions onto the $M^2_{\jp \pi}$ axis are shown
in Fig.~\ref{fig:fitresdef_contrib}.
Projections onto the angular variables for the region defined by
$\sx>1.2\,\gevccsq$, $16\,\gevccsq<\sy<19\,\gevccsq$ (the intersection of the
second horizontal slice and the second, third and fourth vertical slices,
where the $\zp{4200}$ signal is mostly concentrated)
are shown in Fig.~\ref{fig:fitresang}.
A comparison of the fit results with and without the $\zp{4430}$
with the $\zp{4200}$ included in the model is shown in Fig.~\ref{fig:z4430_2}. 

\begin{table*}
\caption{The absolute values and phases of the helicity amplitudes
in the default model for the $1^+$ spin-parity of the $\zp{4200}$.
Errors are statistical only.}
\begin{tabular}{c|c|c|c|c|c|c}
\hline\hline
Resonance &
$|H_0|$ & $\arg H_0$ & $|H_1|$ & $\arg H_1$ & $|H_{-1}|$ & $\arg H_{-1}$ \\
\hline
$K^*_0(800)$ & $1.12\pm0.04$ & $2.30\pm0.04$ &
                --- & --- & --- & --- \\
$K^*(892)$ & $1.0$ (fixed) & $0.0$ (fixed) &
              $(8.44\pm0.10)\times10^{-1}$ & $3.14\pm0.03$ &
              $(1.96\pm0.14)\times10^{-1}$ & $-1.70\pm0.07$ \\
$K^*(1410)$ & $(1.19\pm0.27)\times10^{-1}$ & $0.81\pm0.26$ &
               $(1.23\pm0.38)\times10^{-1}$ & $-1.04\pm0.26$ &
               $(0.36\pm0.39)\times10^{-1}$ & $0.67\pm1.06$ \\
$K^*_0(1430)$ & $(8.90\pm0.28)\times10^{-1}$ & $-2.17\pm0.05$ &
                 --- & --- & --- & --- \\
$K^*_2(1430)$ & $4.66\pm0.18$ & $-0.32\pm0.05$ &
                 $4.65\pm0.18$ & $-3.05\pm0.08$ &
                 $1.26\pm0.23$ & $-1.92\pm0.20$ \\
$K^*(1680)$ & $(1.39\pm0.43)\times10^{-1}$ & $-2.46\pm0.31$ &
               $(0.82\pm0.48)\times10^{-1}$ & $-2.85\pm0.49$ &
               $(1.61\pm0.56)\times10^{-1}$ & $1.88\pm0.28$ \\
$K^*_3(1780)$ & $16.8\pm3.6$ & $-1.43\pm0.24$ &
                 $19.1\pm4.5$ & $2.03\pm0.31$ &
                 $10.2\pm5.2$ & $1.55\pm0.62$ \\
$K^*_0(1950)$ & $(2.41\pm0.60)\times10^{-1}$ & $-2.39\pm0.25$ &
                 --- & --- & --- & --- \\
$K^*_2(1980)$ & $4.53\pm0.74$ & $-0.26\pm0.16$ &
                 $3.78\pm0.98$ & $3.08\pm0.28$ &
                 $3.51\pm1.03$ & $2.63\pm0.34$ \\
$K^*_4(2045)$ & $590\pm136$ & $-2.66\pm0.23$ &
                 $676\pm164$ & $0.06\pm0.25$ &
                 $103\pm174$ & $-1.03\pm1.62$ \\
$\zp{4430}$   & $1.12\pm0.32$ & $-0.31\pm0.26$ & $1.17\pm0.46$ & $0.77\pm0.25$ & \multicolumn{2}{c}{$H_{-1}=H_1$} \\
$\zp{4200}$   & $0.71\pm0.37$ & $2.14\pm0.40$ & $3.23\pm0.79$ & $3.00\pm0.15$ & \multicolumn{2}{c}{$H_{-1}=H_1$} \\
\hline\hline
\end{tabular}
\label{tab:defamp}
\end{table*}

\begin{table}[ht]
\caption{
The fit fractions and significances of all resonances in the default
model ($J^P=1^+$).
}
\begin{center}
\begin{tabular}{c|c|c}
\hline\hline
Resonance & Fit fraction & Significance (Wilks) \\
\hline
$K_0^*(800)$ & $(7.1^{+0.7}_{-0.5})\%$  & $22.5\sigma$ \\
$K^*(892)$   & $(69.0^{+0.6}_{-0.5})\%$ & $166.4\sigma$ \\
$K^*(1410)$  & $(0.3^{+0.2}_{-0.1})\%$  & $4.1\sigma$ \\
$K_0^*(1430)$& $(5.9^{+0.6}_{-0.4})\%$  & $22.0\sigma$ \\
$K_2^*(1430)$& $(6.3^{+0.3}_{-0.4})\%$  & $23.5\sigma$ \\
$K^*(1680)$  & $(0.3^{+0.2}_{-0.1})\%$  & $2.7\sigma$ \\
$K_3^*(1780)$& $(0.2^{+0.1}_{-0.1})\%$  & $3.8\sigma$ \\
$K_0^*(1950)$& $(0.1^{+0.1}_{-0.1})\%$  & $1.2\sigma$ \\
$K_2^*(1980)$& $(0.4^{+0.1}_{-0.1})\%$  & $5.3\sigma$ \\
$K_4^*(2045)$& $(0.2^{+0.1}_{-0.1})\%$  & $3.8\sigma$ \\
$\zp{4430}$  & $(0.5^{+0.4}_{-0.1})\%$        & $5.1\sigma$ \\
$\zp{4200}$  & $(1.9^{+0.7}_{-0.5})\%$  & $8.2\sigma$ \\
\hline\hline
\end{tabular}
\end{center}
\label{tab:ffrac}
\end{table}

\begin{figure*}
\includegraphics[width=6cm]{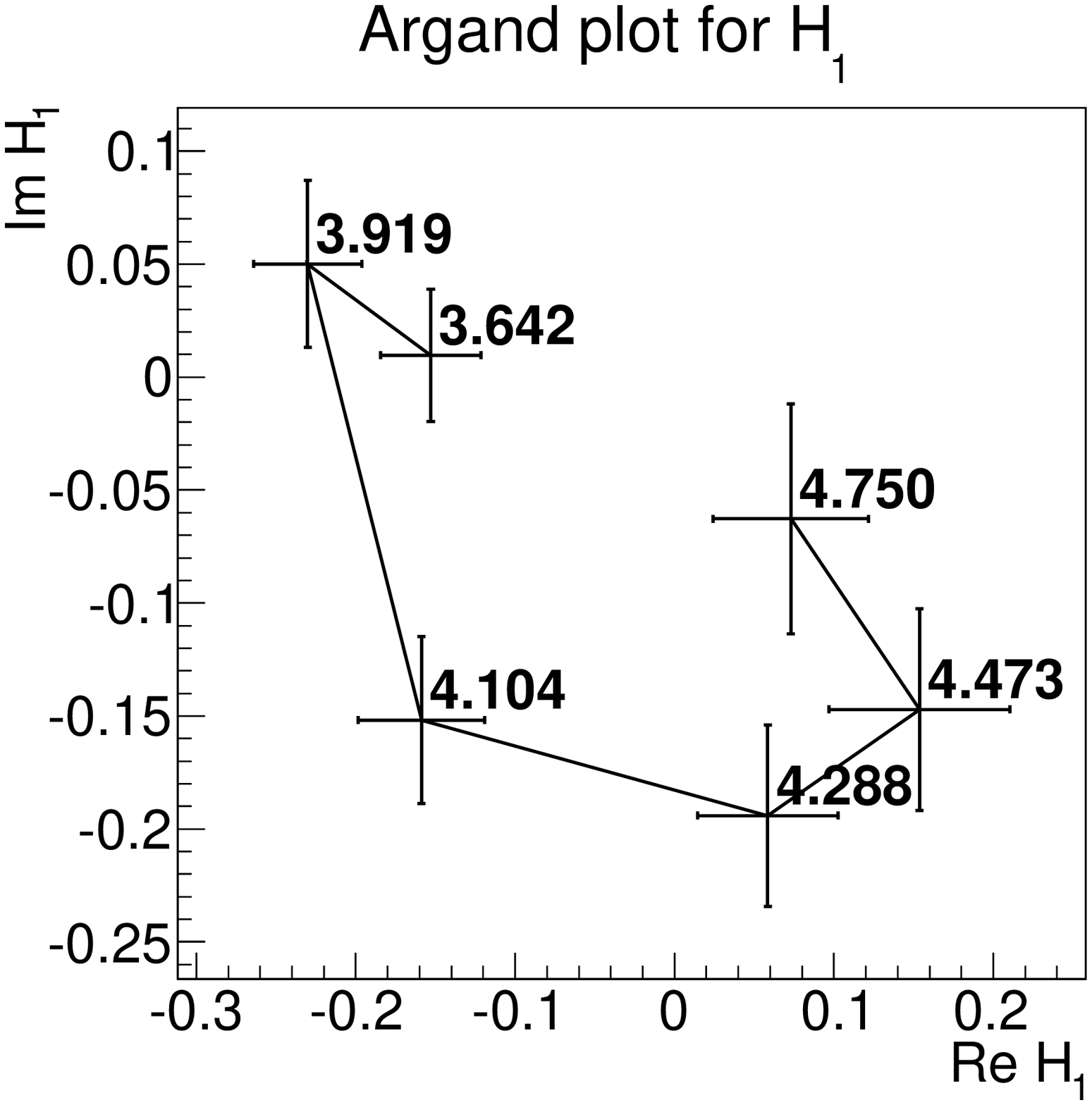}
\includegraphics[width=6cm]{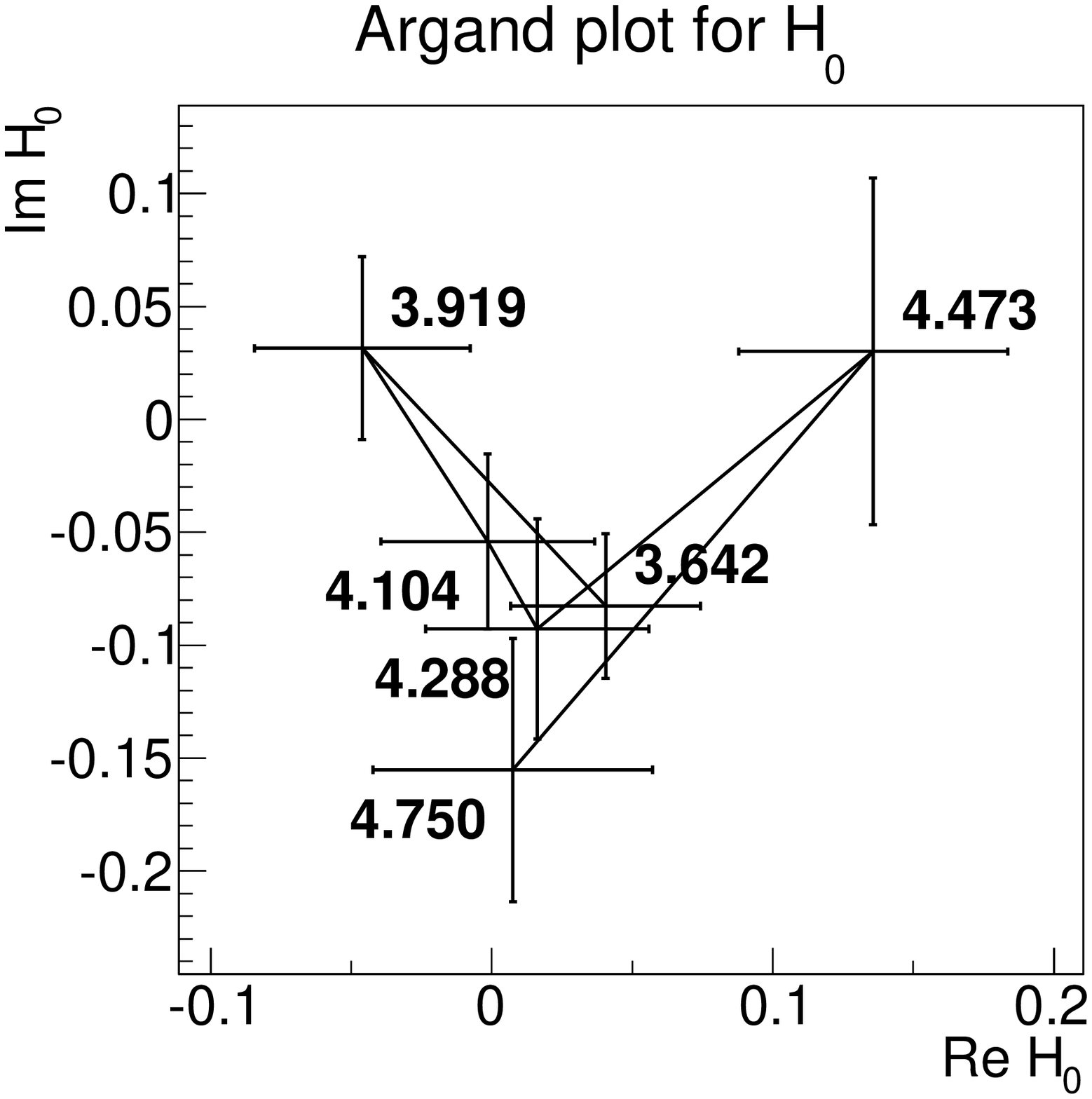}
\caption{Argand plots for the $\zp{4200}$ helicity amplitudes. The bin central
mass values (in $\gevcc$) are shown near the points.}
\label{fig:argand}
\end{figure*}

We also perform a fit with the $\zp{4200}$ Breit-Wigner amplitude
changed to a combination of constant amplitudes. We use 6 bins with
borders at $M_0 - 2 \Gamma_0$, $M_0 - \Gamma_0$, $M_0 - 0.5 \Gamma_0$,
$M_0$, $M_0 + 0.5 \Gamma_0$, $M_0 + \Gamma_0$ and $M_0 + 2 \Gamma_0$,
where $M_0$ and $\Gamma_0$ are the fit results for the mass and width
of the $\zp{4200}$ in the default model. We use two independent sets of
constant amplitudes to represent the two helicity amplitudes of the $\zp{4200}$,
$H_0$ and $H_1$. The two sets of amplitudes are measured simultaneously.
The results are shown in Fig.~\ref{fig:argand}. The Argand
plot for $H_1$ clearly shows a resonancelike change of the amplitude absolute
value and phase. Because the Argand plot for the $H_0$ amplitudes has much
larger relative errors, it is not possible to draw any conclusions from it.

We check if the $\zp{4200}$ signal can be explained by a resonance in the
$\jp \km$ system by adding a $\jp \km$ resonance, which is referred to as
the $\zs$ instead of the $\zp{4200}$. The preferred quantum numbers of the
$\zs$ are also $J^P=1^+$; the mass and width in the default model for the $1^+$
hypothesis are $4228\pm5\ \mevcc$ and $30\pm17\ \mev$, respectively.
The Wilks significance is only $4.3\sigma$. The hypothesis of the existence of
a $\jp \pip$ resonance is preferred over the hypothesis of the existence of
a $\jp \km$ resonance at the level of $7.4 \sigma$. The $\zs$ becomes
insignificant if the $\zp{4200}$ is added to the model.

Separate results from $\jp\to\elp\elm$ and $\jp\to\mup\mum$ decay samples
agree with each other and with the results from the combined sample.
The $\zp{4200}$ mass, width and significance for the $J^P=1^+$ hypothesis
for each $\jp$ decay channel are shown in Table~\ref{tab:eemumu}.

\begin{table}
\caption{Comparison of the $\zp{4200}$ parameters in the decay channels
$\jp\to\elp\elm$, $\jp\to\mup\mum$ and the combined sample in the default model
($J^P=1^+$).}
\begin{tabular}{c|c|c|c}
\hline\hline
Sample & combined & $\jp\to\elp\elm$ & $\jp\to\mup\mum$ \\
\hline
Mass, $\mevcc$ & $4196^{+31}_{-29}$ & $4198\pm41$ & $4217\pm41$ \\
Width, $\mev$ & $370\pm70$ & $358\pm57$ & $443\pm94$ \\
Significance (Wilks) & $8.2\sigma$ & $5.4\sigma$ & $5.3\sigma$ \\
\hline\hline
\end{tabular}
\label{tab:eemumu}
\end{table}

We also consider other amplitude models: without one
of the insignificant $\kst$ resonances
[$K^*(1680)$, $K^*_0(1950)$];
with the addition of S-, P- and D-wave nonresonant $\km\pip$ amplitudes;
with free Blatt-Weisskopf $r$ parameters;
with free masses and widths of $K^*$ resonances
(with Gaussian constraints to their known values~\cite{PDG}) and with the LASS
amplitudes~\cite{LASS}
instead of Breit-Wigner amplitudes for all spin-0 $K^*$ resonances.

The significances of the $\zp{4200}$ for all models other than the default
are shown in Table~\ref{tab:zsigsyst}.
The minimal Wilks significance for the $1^+$ hypotheses is
$6.6\sigma$; the corresponding global significance is $6.2\sigma$.

\begin{table}
\caption{Model dependence of the $\zp{4200}$ Wilks significance.}
\begin{tabular}{c|c|c|c|c|c}
\hline\hline
Model & $0^-$ & $1^-$ & $1^+$ & $2^-$ & $2^+$ \\
\hline
Without $K^*(1680)$        &
$3.2\sigma$ & $3.1\sigma$ & $8.4\sigma$ & $3.7\sigma$ & $1.9\sigma$ \\
Without $K^*_0(1950)$        &
$3.6\sigma$ & $2.8\sigma$ & $8.6\sigma$ & $5.0\sigma$ & $2.6\sigma$ \\
LASS                       &
$3.8\sigma$ & $1.0\sigma$ & $6.6\sigma$ & $5.2\sigma$ & $2.3\sigma$ \\
Free masses and widths     &
$2.4\sigma$ & $1.6\sigma$ & $7.3\sigma$ & $4.6\sigma$ & $1.9\sigma$ \\
Free $r$                   &
$5.0\sigma$ & $2.6\sigma$ & $8.4\sigma$ & $4.5\sigma$ & $0.9\sigma$ \\
Nonresonant ampl. (S)     &
$3.8\sigma$ & $2.9\sigma$ & $7.9\sigma$ & $4.1\sigma$ & $2.0\sigma$ \\
Nonresonant ampl. (S,P)   &
$3.7\sigma$ & $2.4\sigma$ & $7.7\sigma$ & $3.7\sigma$ & $1.4\sigma$ \\
Nonresonant ampl. (S,P,D) &
$4.1\sigma$ & $2.3\sigma$ & $7.7\sigma$ & $3.8\sigma$ & $1.3\sigma$ \\
\hline\hline
\end{tabular}
\label{tab:zsigsyst}
\end{table}

The exclusion levels of the spin-parity hypotheses ($J^P=j^p$,
$j^p \in \{0^+,\,1^-,\,2^-,\,2^+\}$)
for the default model are calculated using MC simulation. The procedure
is the same as in Ref.~\cite{z4430jp}.
We generate MC pseudoexperiments in accordance with the fit result
with the $j^p$ $\zp{4200}$ signal in data and fit them with
the $j^p$ and $1^+$ signals.
The resulting distribution of
$\dlnl = (-2\ln L)_{J^P=j^p} - (-2\ln L)_{J^P=1^+}$
is fitted to an asymmetric Gaussian function and the $p$-value
is calculated as the integral of the fitting function normalized to 1
from the value of
$\dlnl$ in data to $+\infty$. The results are presented in Table
~\ref{tab:excl}.

\begin{figure}
\includegraphics[width=6cm]{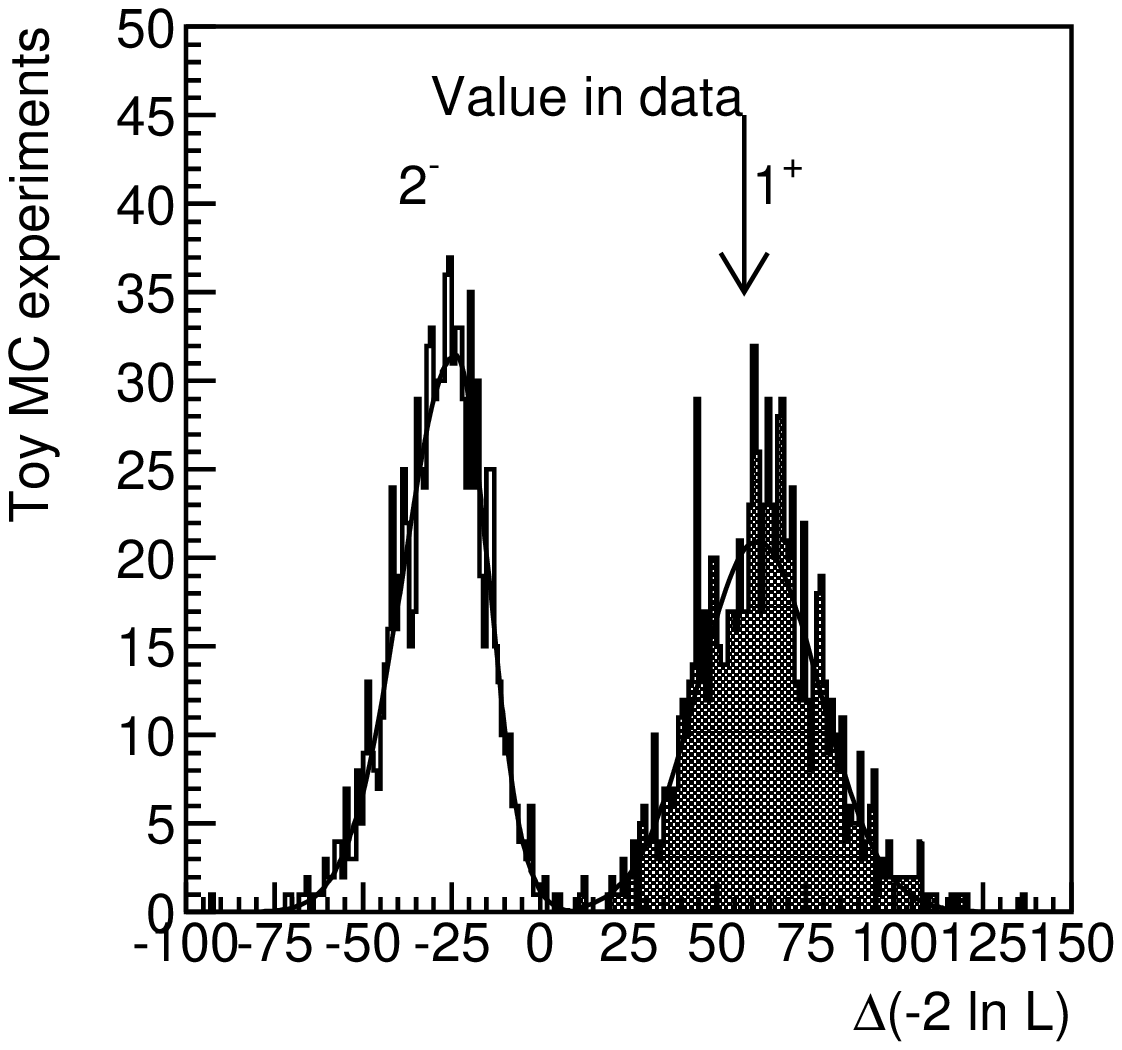}
\caption{Comparison of the $2^-$ and $1^+$ hypotheses in the default model.
The histograms are distributions of $\dlnl$ in MC pseudoexperiments
generated in accordance with the fit results with $2^-$
(open histogram) and $1^+$ (hatched histogram) $\z$ signals.
The $\dlnl$ value observed in data is indicated with an arrow.}
\label{fig:toy}
\end{figure}

We also generate MC pseudoexperiments in accordance with the fit results for
the $1^+$ hypothesis, fit them with the $j^p$ and $1^+$ signals
 and obtain the distribution of $\dlnl$.
This distribution is fitted to an asymmetric Gaussian function and the
confidence level of the $1^+$ hypothesis is calculated as the integral
of the fitting function
normalized to 1 from $-\infty$ to the value of $\dlnl$ in data.
The resulting confidence levels are shown in Table~\ref{tab:excl}.
The distributions of $\dlnl$ for $j^p=2^-$
are shown in Fig.~\ref{fig:toy}.

For models other than the default, we do not use the calculation of
exclusion levels of the spin-parity hypotheses based on MC pseudoexperiments.
Instead, the significance
of the $1^+$ hypothesis over the $j^p$ hypothesis is estimated as
$\sqrt{\dlnl}$. The comparison of the two methods for the default model
is shown in Table~\ref{tab:excl}. The formula-based calculation results
in smaller values of the significance than the MC-based calculation and, thus,
it provides a conservative estimate of the significance. The results for
all models are shown in Table~\ref{tab:exclsyst}.
The $1^+$ hypothesis is favored over the $0^-$, $1^-$, $2^-$, $2^+$
hypotheses at the levels of $6.1\sigma$,
$7.4\sigma$, $4.4\sigma$ and $7.0\sigma$, respectively.

\begin{table}
\caption{Exclusion levels of the $\zp{4200}$ spin-parity hypotheses and
confidence levels of the $1^+$ hypothesis for the default model.}
\begin{tabular}{c|c|c|c}
\hline\hline
\multirow{2}{*}{$j^p$} & \multicolumn{2}{|c|}{$1^+$ over $j^p$} &
\multirow{2}{*}{$1^+$ C. L.} \\
\cline{2-3}
& MC & $\sqrt{\dlnl}$ & \\
\hline
$0^-$ & $8.6\sigma$ & $7.9\sigma$ & 26\% \\
$1^-$ & $9.8\sigma$ & $8.7\sigma$ & 48\% \\
$2^-$ & $8.8\sigma$ & $7.6\sigma$ & 40\% \\
$2^+$ & $10.6\sigma$ & $8.8\sigma$ & 42\% \\
\hline\hline
\end{tabular}
\label{tab:excl}
\end{table}

\begin{table}
\caption{Exclusion levels of the $\zp{4200}$ spin-parity hypotheses.}
\begin{tabular}{c|c|c|c|c}
\hline\hline
Model & $0^-$ & $1^-$ & $2^-$ & $2^+$ \\
\hline
Without $K^*(1680)$       & $8.5\sigma$ & $8.5\sigma$ & $8.0\sigma$ & $9.0\sigma$ \\
Without $K^*_0(1950)$     & $8.4\sigma$ & $8.8\sigma$ & $7.3\sigma$ & $8.9\sigma$ \\
LASS                      & $6.1\sigma$ & $7.4\sigma$ & $4.4\sigma$ & $7.0\sigma$ \\
Free masses and widths    & $7.6\sigma$ & $7.9\sigma$ & $5.9\sigma$ & $7.8\sigma$ \\
Free $r$                  & $7.4\sigma$ & $8.7\sigma$ & $7.5\sigma$ & $9.2\sigma$ \\
Nonresonant ampl. (S)     & $7.6\sigma$ & $8.1\sigma$ & $7.2\sigma$ & $8.5\sigma$ \\
Nonresonant ampl. (S,P)   & $7.4\sigma$ & $8.1\sigma$ & $7.2\sigma$ & $8.4\sigma$ \\
Nonresonant ampl. (S,P,D) & $7.2\sigma$ & $8.1\sigma$ & $7.1\sigma$ & $8.4\sigma$ \\
\hline\hline
\end{tabular}
\label{tab:exclsyst}
\end{table}

The results of the study of the model dependence of the $\zp{4200}$
mass and width are shown in Table~\ref{tab:syst}.
The maximal deviations of the mass and the width of the $\zp{4200}$
from the default model values are considered as the systematic uncertainty
due to the amplitude model dependence.

We also estimate the systematic error associated with
the uncertainties in the modeling of the background distribution by varying
the background parameters by $\pm1\sigma$ (with other parameters varied in
accordance with the correlation coefficients)
and performing the fit to the data.
The maximal deviations are considered as the systematic error due
to the background parameterization uncertainty. This error is found to be
negligibly small compared to the error
due to amplitude model dependence
for all the results.

\begin{table}
\caption{Systematic uncertainties in the $\zp{4200}$ mass (in $\mevcc$)
and width (in $\mev$).}
\begin{tabular}{c|c|c}
\hline\hline
Model or error source & $\ $Mass $\ $ & Width \\
\hline
Without $K^*(1680)$        &
$^{+0}_{-1}$  & $^{+0}_{-34}$ \\
Without $K^*_0(1950)$        &
$^{+9}_{-0}$  & $^{+0}_{-54}$ \\
LASS                       &
$^{+0}_{-13}$  & $^{+0}_{-132}$ \\
Free masses and widths     &
$^{+0}_{-3}$  & $^{+0}_{-29}$  \\
Free $r$                   &
$^{+16}_{-0}$ & $^{+58}_{-0}$ \\
Nonresonant ampl. (S)     &
$^{+0}_{-6}$  & $^{+0}_{-15}$  \\
Nonresonant ampl. (S,P)   &
$^{+17}_{-0}$  & $^{+70}_{-0}$  \\
Nonresonant ampl. (S,P,D) &
$^{+0}_{-0}$  & $^{+6}_{-0}$  \\
\hline
Amplitude model, total &
$^{+17}_{-13}$ & $^{+70}_{-132}$ \\
\hline\hline
\end{tabular}
\label{tab:syst}
\end{table}

Using the helicity amplitudes shown in Table~\ref{tab:defamp}, one
can calculate the amplitudes in the transversity basis:
\begin{equation}
A_0 = H_0,\ \atang = \frac{H_1 + H_{-1}}{\sqrt{2}},
\ \anorm = \frac{H_1 - H_{-1}}{\sqrt{2}},
\label{eq:heltra}
\end{equation}
where $A_0$, $\atang$ and $\anorm$ are the transversity amplitudes.
The amplitudes from Table~\ref{tab:defamp} should be normalized so that,
for a $\kst$ resonance,
\begin{equation}
|H_0|^2 + |H_1|^2 + |H_{-1}|^2 = 1
\end{equation}
before the application of Eq.~\eqref{eq:heltra}. The resulting transversity
amplitudes for the $\kst(892)$ are shown in Table~\ref{tab:ksttra}.
The transversity amplitude systematic errors are
due to amplitude model dependence.
The results agree with previous Belle measurements
for the $(B^0 + \bar{B}^0)$ sample in Ref.~\cite{belletransversity}
and supersede them.

\begin{table}[ht]
\caption{The transversity amplitudes of the $\kst(892)$.}
\begin{tabular}{c|c}
\hline\hline
Parameter & Result \\
\hline
$|\atang|^2$  & $0.227\pm0.007\pm0.006$ \\
$|\anorm|^2$  & $0.201\pm0.007\pm0.005$ \\
$\arg \atang$ & $-2.92\pm0.04\pm0.04$ \\
$\arg \anorm$ & $2.91\pm0.03\pm0.03$  \\
\hline\hline
\end{tabular}
\label{tab:ksttra}
\end{table}

We perform a search for
the $\zp{3900}$, using the amplitude model with
the $\zp{4200}$ ($J^P=1^+$) as a null hypothesis.
All quantum number hypotheses with $J \le 2$ are considered
($J^P \in \{0^+,\,1^-,\,1^+,\,2^-\,\text{and}\,2^+\}$).
The mass and the width of the $\zp{3900}$ are constrained
in accordance with Eq.~\eqref{eq:z4430masslim}.
The average result of BESIII~\cite{z3900bes},
Belle~\cite{z3900belle} and analysis based on CLEO data~\cite{z3900cleo},
\begin{equation*}
M_0 = 3891.2\pm3.3\ \mevcc,\ \Gamma_0 = 39.5\pm8.1\ \mev,
\end{equation*}
is used as the nominal mass and width of the $\zp{3900}$.
The results are shown in Table~\ref{tab:z3900z4430}. No significant
signal is found.
\begin{table*}[ht]
\caption{Fit results with addition of the $\zp{3900}$ in the default model.
Errors are statistical only.}
\begin{tabular}{c|c|c|c|c|c}
\hline\hline
$J^P$ & $0^-$ & $1^-$ & $1^+$ & $2^-$ & $2^+$ \\
\hline
Mass, $\mevcc$ & $3889.8\pm3.3$ & $3890.3\pm3.1$ & $3890.6\pm3.3$ & $3891.1\pm3.2$ & $3891.5\pm3.3$ \\
Width, $\mev$ & $43.2\pm6.5$ & $37.8\pm7.9$ & $39.2\pm8.1$ & $39.4\pm8.5$ & $41.2\pm7.7$ \\
Significance & $2.4\sigma$ & $1.1\sigma$ & $0.1\sigma$ & $<0.1\sigma$ & $0.2\sigma$ \\
\hline\hline
\end{tabular}
\label{tab:z3900z4430}
\end{table*}

\subsection{Efficiency and branching fractions}

We use the signal density function determined from the fits to
calculate the efficiency
\begin{equation}
\epsilon_0 = \frac
{\int S(\Phi) \epsilon(\Phi) d\Phi}
{\int S(\Phi) d\Phi},
\end{equation}
where $\epsilon(\Phi)$ is the phase-space-dependent efficiency.
The ratio of integrals is calculated with a
the Monte-Carlo method without efficiency parameterization.
The reconstruction efficiency is found to be $\epsilon_0 = (28.4\pm1.1)\%$.
The central value is calculated for the default model with $\z$ ($J^P=1^+$).
The efficiency includes the correction for the difference between the particle
identification efficiency in MC and data, $(93.1\pm3.5)\%$.
The relative error of the efficiency includes
the uncertainty in track reconstruction efficiency ($1.4\%$),
the error from the particle identification efficiency difference
between MC and data ($3.8\%$) and the uncertainty due to the amplitude model
dependence ($0.3\%$).
The error due to MC statistics is negligibly small.

Using the obtained efficiency and the branching fractions for $\jp$ decays
to $\ee$ and $\mumu$~\cite{PDG}, we determine:
\begin{equation*}
\br(\decay) = (1.15\pm0.01\pm0.05)\times10^{-3}.
\end{equation*}
This result assumes equal production of $B^0\bar{B}^0$ and $B^+B^-$ pairs.
The central value is given for the default model with the $J^P=1^+$
assignment for the $\zp{4200}$.
The systematic error includes the uncertainty in the efficiency,
the number of $B$ mesons (1.4\%), the signal yield (0.3\%)
and the $\jp\to\lp\lm$ branching fraction (1.0\%).

The fit fraction of a resonance $R$ [the $\zp{4200}$, $\zp{4430}$ or one of the
$K^*$ resonances]
is defined as
\begin{equation}
f = \frac{\int S_R(\Phi) d\Phi}
{\int S(\Phi) d\Phi},
\end{equation}
where $S_R(\Phi)$ is the signal density function with all contributions other than
the contribution of the $R$ resonance set to 0.
The statistical uncertainties in the fit fractions are determined
from a set of MC pseudoexperiments generated in accordance with the
fit result in data.
We fit each sample and calculate the fit fractions;
the resulting distribution of the fit fractions is fitted to
an asymmetric Gaussian function with peak position fixed at the fit fraction
in data.
The standard deviations of the Gaussian function are treated as the statistical
uncertainties. We find good agreement between the distributions of the
fit fractions in the pseudoexperiments with the fitting function for
all resonances except for the $K^*(892)$. For the $K^*(892)$, we release
the peak
position and treat the difference between the resulting fit fraction and the
fit fraction in data (-0.42\% absolute or -0.61\% relative) as an additional
systematic error due to fit bias.
The results are summarized in Table~\ref{tab:ffrac}. 

The branching fraction of $B^0\to\jp\kst(892)$ decay is given by
\begin{equation}
\begin{split}
&\br(\bar{B}^0\to\jp\kst(892)) = \\
&\qquad 1.5\,f_{\kst(892)}\,\br(\decay), \\
\end{split}
\end{equation}
where $f_{\kst(892)}$ is the fit fraction of the $\kst(892)$. The result
is
\begin{equation*}
\br(\bar{B}^0\to\jp\kst(892)) = (1.19\pm0.01\pm0.08)\times10^{-3}.
\end{equation*}
The systematic error includes contributions from the same sources
as the uncertainty in
the branching fraction of $\decay$ decay,
fit bias for the $\kst(892)$ fit fraction (-0.6\%)
and
the amplitude model [$(^{+1.5}_{-2.0})\%$] dependence
of the $\kst(892)$ fit fraction.

The branching fraction products for the $\zp{4430}$ and $\zp{4200}$ are
\begin{equation*}
\begin{split}
&\br(\bar{B}^0\to\zp{4430}\km)\times\br(\zp{4430}\to\jp\pip) = \\
&\quad(5.4^{+4.0 +1.1}_{-1.0 -0.9})\times10^{-6}, \\
&\br(\bar{B}^0\to\zp{4200}\km)\times\br(\zp{4200}\to\jp\pip) = \\
&\quad(2.2^{+0.7 +1.1}_{-0.5 -0.6})\times10^{-5}, \\
\end{split}
\end{equation*}
where the systematic error due to the amplitude model dependence is
$(^{+19.9}_{-14.9})\%$ and $(^{+49.0}_{-26.7})\%$, respectively.

In the determination of the product of branching fractions for
the $\zp{3900}$, its quantum numbers are assumed to be $J^P=1^+$
in accordance with the result of the BESIII angular analysis
of the $D\bar{D}^*$ decay mode~\cite{z3900bes_ddst}.
The result is
\begin{equation*}
\begin{split}
&\br(\bar{B}^0\to\zp{3900}\km)\times\br(\zp{3900}\to\jp\pip) < \\
&\quad9 \times 10^{-7}\ \text{(90\% CL)}. \\
\end{split}
\end{equation*}


\section{Conclusions}

An amplitude analysis of $\decay$ decays in four dimensions
has been performed.
A new charged charmoniumlike state $\zp{4200}$ decaying to $\jp$ and $\pip$
is observed with the significance of $6.2\sigma$.
The minimal quark content of this state is exotic:
$|c\bar{c}u\bar{d}\rangle$.
Its mass and width are measured to be
\begin{equation*}
\begin{split}
&M = 4196^{+31 +17}_{-29 -13}\ \mevcc, \\
&\Gamma = 370^{+70 +70}_{-70 -132}\ \mev.
\end{split}
\end{equation*}
The preferred quantum number assignment is $J^P=1^+$. Other hypotheses with
$J^P \in \{0^-,\,1^-,\,2^-,\,2^+\}$ are excluded at the levels of $6.1\sigma$,
$7.4\sigma$, $4.4\sigma$ and $7.0\sigma$, respectively.
Also, evidence for a new decay channel $\to\jp\pip$ of the $\zp{4430}$ is
found.

The LHCb Collaboration included a second $\z$ state in the
amplitude analysis of $\bar{B}^0\to\psp\km\pip$ decays
together with the $\zp{4430}$, but did not claim an
observation~\cite{z4430lhcb}.
The reported mass and width of this second $\z$ are close to the mass and
width of the $\zp{4200}$ and, while
the preferred quantum number assignment of the quantum numbers
is $J^P=0^-$, $J^P=1^+$ is not excluded.
Thus, the effect observed in Ref.~\cite{z4430lhcb} may be due to
$\zp{4200}\to\psp\pip$.

The branching fractions are found to be
\begin{equation*}
\begin{split}
&\br(\decay) = (1.15\pm0.01\pm0.05)\times10^{-3}, \\
&\br(\bar{B}^0\to\jp\kst(892)) = (1.19\pm0.01\pm0.08)\times10^{-3}, \\
&\br(\bar{B}^0\to\zp{4430}\km)\times\br(\zp{4430}\to\jp\pip) = \\
&\quad(5.4^{+4.0 +1.1}_{-1.0 -0.9})\times10^{-6}, \\
&\br(\bar{B}^0\to\zp{4200}\km)\times\br(\zp{4200}\to\jp\pip) = \\
&\quad(2.2^{+0.7 +1.1}_{-0.5 -0.6})\times10^{-5}, \\
&\br(\bar{B}^0\to\zp{3900}\km)\times\br(\zp{3900}\to\jp\pip) < \\
&\quad9 \times 10^{-7}\ \text{(90\% CL)}. \\
\end{split}
\end{equation*}


\section{Acknowledgments}

We thank the KEKB group for the excellent operation of the
accelerator; the KEK cryogenics group for the efficient
operation of the solenoid; and the KEK computer group,
the National Institute of Informatics, and the 
PNNL/EMSL computing group for valuable computing
and SINET4 network support.  We acknowledge support from
the Ministry of Education, Culture, Sports, Science, and
Technology (MEXT) of Japan, the Japan Society for the 
Promotion of Science (JSPS), and the Tau-Lepton Physics 
Research Center of Nagoya University; 
the Australian Research Council and the Australian 
Department of Industry, Innovation, Science and Research;
Austrian Science Fund under Grant No.~P 22742-N16 and P 26794-N20;
the National Natural Science Foundation of China under Contracts 
No.~10575109, No.~10775142, No.~10875115, No.~11175187, and  No.~11475187; 
the Ministry of Education, Youth and Sports of the Czech
Republic under Contract No.~LG14034;
the Carl Zeiss Foundation, the Deutsche Forschungsgemeinschaft
and the VolkswagenStiftung;
the Department of Science and Technology of India; 
the Istituto Nazionale di Fisica Nucleare of Italy; 
National Research Foundation (NRF) of Korea Grants
No.~2011-0029457, No.~2012-0008143, No.~2012R1A1A2008330, 
No.~2013R1A1A3007772, No.~2014R1A2A2A01005286, No.~2014R1A2A2A01002734, 
No.~2014R1A1A2006456;
the Basic Research Lab program under NRF Grant No.~KRF-2011-0020333, 
No.~KRF-2011-0021196, Center for Korean J-PARC Users,
No.~NRF-2013K1A3A7A06056592; 
the Brain Korea 21-Plus program and the Global Science Experimental Data 
Hub Center of the Korea Institute of Science and Technology Information;
the Polish Ministry of Science and Higher Education and 
the National Science Center;
the Ministry of Education and Science of the Russian
Federation (particularly under Contract No.~14.A12.31.0006),
the Russian Federal Agency for Atomic Energy
and the Russian Foundation for Basic Research under Grant No.~14-02-01220;
the Slovenian Research Agency;
the Basque Foundation for Science (IKERBASQUE) and 
the Euskal Herriko Unibertsitatea (UPV/EHU) under program UFI 11/55 (Spain);
the Swiss National Science Foundation; the National Science Council
and the Ministry of Education of Taiwan; and the U.S.\
Department of Energy and the National Science Foundation.
This work is supported by a Grant-in-Aid from MEXT for 
Science Research in a Priority Area (``New Development of 
Flavor Physics'') and from JSPS for Creative Scientific 
Research (``Evolution of Tau-lepton Physics'').

\appendix
\section{Calculation of the local, Wilks and global significance}
\label{sec:significance}

\begin{figure}
\includegraphics[width=6cm]{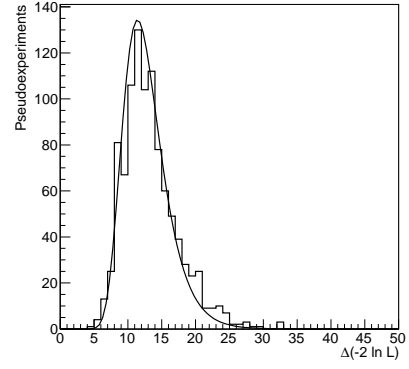}
\caption{Results of a fit to a $\dlnl$ distribution done as part of the 
$\zp{4200}$ global significance calculation for the $J^P=1^+$ hypothesis.}
\label{fig:lee}
\end{figure}

For the significance calculation, one needs to know the distribution of the
difference between the $-2 \ln L$ values with and without a $\z$ contribution
provided that there is no $\z$ signal.
The Wilks significance is given by Wilks' theorem~\cite{wilks}:
\begin{equation}
p(\delta) = \int\limits_\delta^{+\infty}\chi^2_\kappa(x) dx =
\frac{\Gamma(\frac{\kappa}{2},\delta/2)}{\Gamma(\frac{\kappa}{2})},
\label{eq:taildf_wilks}
\end{equation}
where $p(\delta)$ is the probability that $\dlnl > \delta$;
$\kappa$ is the number of degrees of freedom of the $\chi^2$ distribution,
which is equal to the number of additional free parameters
and $\Gamma(\kappa/2,\delta/2)$ is the upper incomplete gamma function
[$\Gamma(a,x)=\int_x^{+\infty} t^{a-1} e^{-t} dt$].
In this analysis, the number of additional free parameters is four for
$J^P$ = $0^-$, $1^-$ and $2^+$ or six for $J^P$ = $1^+$ and $2^-$. These
parameters include mass, width and one or two complex amplitudes. The local
significance is the significance with fixed mass and width; it is
given by Eq.~\eqref{eq:taildf_wilks} with $\kappa \to \kappa - 2$.

The mass and the width of the resonance are defined only under an alternative
(\textit{i.e.,} when amplitudes are not equal to 0), thus the real distribution
of $\dlnl$ may deviate from the prediction of Wilks' theorem. Furthermore,
since the search is performed over two variables, the one-dimensional
upcrossing method~\cite{gross_vitells_1d} is not valid.

For large values of $\delta$, the $p$-value is the same as the expectation
of the Euler characteristic of the excursion set~\cite{gross_vitells_nd}.
This expectation $E(\delta)$
is given in Ref.~\cite{adler_taylor} [Eq.~(15.10.1) and
Theorem~15.10.1]; it has the form
\begin{equation}
E(\delta) = \sum\limits_{j = 0}^{n-k}
\sum\limits_{l=0}^{\lfloor\frac{j-1}{2}\rfloor}
\sum\limits_{m = 0}^{j - 1 - 2l}
C_{jml} \delta^{(k - j)/2 + m + l} e^{-\delta/2},
\end{equation}
where $n-k$ is the dimension, $k$ is the number of degrees of
freedom (for the application in question, $n$ is the total number of additional
free parameters and $n-k$ is equal to the number of additional free parameters
defined only under alternative),
$\lfloor\,\rfloor$ is the floor function (the largest integer that is not
greater than the argument) and $C_{jml}$ are
constants. The contribution with the largest power of $\delta$
corresponds to $m = j - 1$, $l = 0$, $j = n - k$:
\begin{equation}
E(\delta) \propto \delta^{\frac{n}{2}-1} e^{-\delta/2}.
\label{eq:pval}
\end{equation}

In Ref.~\cite{z4430lhcb}, the global significance is calculated by
fitting the distribution of $\dlnl$
to the $\chi^2_\kappa$ distribution with variable number of degrees of
freedom $\kappa$. The $p$-value is then given by Eq.~\eqref{eq:taildf_wilks};
for large $\delta$, it is approximately equal to
\begin{equation}
p(\delta) \approx \frac{(\delta/2)^{\frac{\kappa}{2} - 1} e^{-\delta/2}}
{\Gamma(\frac{\kappa}{2})}.
\end{equation}
This only coincides with the expected tail distribution of $\dlnl$ that
is given by Eq.~\eqref{eq:pval} when $\kappa = n$, \textit{i.e.,} if there is no
look-elsewhere effect.
We follow the general idea of Ref.~\cite{z4430lhcb}
for the calculation of the global significance from the fit to the $\dlnl$
distribution, but construct another probability density function that agrees
with Eq.~\eqref{eq:pval}.

The probability density function is constructed as a generalization
of a particular case of a search of a one-bin peak in a histogram with
$N$ bins with known distribution and normalization.
The $p$-value in a particular
bin is given by Eq.~\eqref{eq:taildf_wilks} with $\kappa=1$.
The $p$-value for the entire histogram is
\begin{equation}
p(\delta) = 1 - \big(1 - \int\limits_\delta^{+\infty}\chi^2_\kappa(x)
dx\big)^N,
\label{eq:pnbins}
\end{equation}
and the corresponding distribution of $\dlnl$, which is
obtained by differentiation of Eq.~\eqref{eq:pnbins}, is
\begin{equation}
f(\Delta) = N \big(1 - \int\limits_\Delta^{+\infty}\chi^2_\kappa(x)
dx\big)^{N - 1}
\chi^2_\kappa(\Delta).
\end{equation}
For large $\Delta$, this is approximately equal to
\begin{equation}
f(\Delta) \approx N \frac{\Delta^{\frac{\kappa}{2}-1}e^{-\Delta/2}}
{2^{\frac{\kappa}{2}}\Gamma(\frac{\kappa}{2})},
\end{equation}
thus
\begin{equation}
p(\delta) \propto \delta^{\frac{\kappa}{2}-1}e^{-\delta/2}.
\end{equation}

If $\kappa$ is equal to the number of additional free parameters $n$, then
Eq.~\eqref{eq:pval} holds for $p(\delta)$.
The distribution of $\dlnl$ is fitted to the function
\begin{equation}
g(\Delta) = C N \big(1 - \int\limits_\Delta^{+\infty}\chi^2_n(x) dx\big)^{N - 1}
\chi^2_n(\Delta).
\label{eq:lee_pdf}
\end{equation}
where $C$ and $N$ are fit parameters. The result for the
search of a $\z$ with $J^P=1^+$ is shown in
Fig.~\ref{fig:lee}; the parameter $N$ is found to be $12.1\pm 0.4$.



\begin{thebibliography}{99}

\bibitem{olsen_godfrey} S.~Godfrey and S.~L.~Olsen,
Ann. Rev. Nucl. Part. Sci. {\bf 58}, 51 (2008).

\bibitem{brambilla}
N.~Brambilla {\it et al.},
Eur.\ Phys.\ J.\ C {\bf 71}, 1534 (2011).

\bibitem{brambilla_qcd_review}
  N.~Brambilla {\it et al.},
  Eur.\ Phys.\ J.\ C {\bf 74}, 2981 (2014).

\bibitem{choiolsen} S.~K.~Choi {\it et al.} (Belle Collaboration),
Phys.\ Rev.\ Lett. {\bf 100}, 142001 (2008).

\bibitem{z4430dalitz} R.~Mizuk {\it et al.} (Belle Collaboration),
Phys.\ Rev.\ D {\bf 80}, 031104(R) (2009).

\bibitem{z4430jp} K. Chilikin {\it et al.} (Belle Collaboration),
Phys.\ Rev.\ D {\bf 88}, 074026 (2013).

\bibitem{mizukchistov} R.~Mizuk {\it et al.} (Belle Collaboration),
Phys.\ Rev.\ D {\bf 78}, 072004 (2008).

\bibitem{babarjpkpi}
B.~Aubert {\it et al.}  (BaBar Collaboration),
Phys.\ Rev.\ D {\bf 79}, 112001 (2009).

\bibitem{babarchickpi} J.~P.~Lees {\it et al.} (BaBar Collaboration),
Phys.\ Rev.\ D {\bf 85}, 052003 (2012).

\bibitem{z4430lhcb} R.~Aaij {\it et al.} (LHCb Collaboration),
Phys.\ Rev.\ Lett.\ {\bf 112}, 222002 (2014).

\bibitem{z3900bes} M.~Ablikim {\it et al.} (BESIII Collaboration),
Phys.\ Rev.\ Lett. {\bf 110}, 252001 (2013).

\bibitem{z3900belle} Z.~Q.~Liu {\it et al.} (Belle Collaboration),
  Phys.\ Rev.\ Lett. {\bf 110}, 252002 (2013).

\bibitem{z3900cleo}
  T.~Xiao, S.~Dobbs, A.~Tomaradze and K.~K.~Seth,
  Phys.\ Lett.\ B {\bf 727}, 366 (2013).

\bibitem{z3900bes_ddst}
  M.~Ablikim {\it et al.} (BESIII Collaboration),
  Phys.\ Rev.\ Lett.\  {\bf 112}, 022001 (2014).

\bibitem{z4020} 
  M.~Ablikim {\it et al.} (BESIII Collaboration),
  Phys.\ Rev.\ Lett. {\bf 111}, 242001, (2013).

\bibitem{bes_dstdst}
  M.~Ablikim {\it et al.} (BESIII Collaboration),
  Phys.\ Rev.\ Lett.\  {\bf 112}, 132001 (2014).

\bibitem{kekb}
{S.~Kurokawa and E.~Kikutani, Nucl. Instrum. Methods Phys. Res. Sect.
 A {\bf 499}, 1 (2003), and other papers included in this Volume;
 T.Abe {\it et al.}, Prog. Theor. Exp. Phys. {\bf 2013}, 03A001 (2013)
 and references therein. }

\bibitem{Belle}
 A.~Abashian {\it et al.} (Belle Collaboration), Nucl. Instrum. Methods
 Phys. Res. Sect. A {\bf 479}, 117 (2002); also see detector section in
 J.~Brodzicka {\it et al.}, Prog. Theor. Exp. Phys. {\bf 2012}, 04D001 (2012).

\bibitem{svd2}
Z.~Natkaniec {\it et al.} (Belle SVD2 Group), Nucl. Instrum. Methods Phys. Res. Sect. A {\bf 560}, 1 (2006).

\bibitem{geant} R.~Brun {\it et al.}, GEANT 3.21, CERN DD/EE/84-1, 1984.

\bibitem{evtgen} D.~J.~Lange, Nucl. Instrum. Methods A {\bf 462}, 152 (2001).

\bibitem{PDG} J.~Beringer {\it et al.} (Particle Data Group),
Phys.\ Rev.\ D {\bf 86}, 010001 (2012).

\bibitem{LASS} D. Aston \textit{et al.}, Nucl. Phys. B \textbf{296}, 493 (1988).

\bibitem{belletransversity} R.~Itoh {\it et al.} (Belle Collaboration),
Phys. Rev. Lett. {\bf 95}, 091601 (2005). 


\bibitem{wilks} S.~S. Wilks, Ann. Math. Statist. {\bf 9}, 60 (1938).

\bibitem{gross_vitells_1d}
  E.~Gross and O.~Vitells,
  Eur.\ Phys.\ J.\ C {\bf 70}, 525 (2010).

\bibitem{gross_vitells_nd}
  O.~Vitells and E.~Gross,
  Astropart.\ Phys.\  {\bf 35}, 230 (2011).

\bibitem{adler_taylor} R.~J.~Adler and J.~E.~Taylor,
{\it Random fields and geometry}, Springer Monographs in Mathematics (2007).
ISBN: 978-0-387-48112-8.

\end{thebibliography}
\end{document}